\documentclass[10pt,reqno]{amsart}

\usepackage{amsmath}
\usepackage{amsthm}
\usepackage{amsfonts}
\usepackage{amssymb}
\usepackage[mathscr]{euscript}
\usepackage{graphicx}
\usepackage[T1]{fontenc}
\usepackage[OT2,T1]{fontenc}
\usepackage[latin1]{inputenc}
\usepackage{lmodern}
\usepackage[all,cmtip]{xy}
\usepackage{cancel}
\usepackage{bbm}
\usepackage[vcentermath]{youngtab}
\usepackage{ stmaryrd }
\usepackage{calc}
\usepackage{color}
\usepackage{setspace}
\usepackage{epigraph}
\usepackage{mathdots}
\usepackage{braket}
\usepackage{longtable}
\usepackage{float}
 \usepackage[foot]{amsaddr}

\usepackage[pdftex,     
          plainpages=false,   
           breaklinks=true,    
           colorlinks=true,
           pdftitle=My Document
           pdfauthor=My Good Self
           colorlinks=true,
	    urlcolor=blue,
	    citecolor=red,
	    linkcolor=blue
          ]{hyperref}
          
\usepackage[a4paper,top=4cm,bottom=4cm,left=2.5cm,right=2.5cm]{geometry}

\theoremstyle{definition}
\newtheorem{defi}{Definition}[section]

\newtheorem{es}{Example}[section]
\theoremstyle{plain}
\newtheorem{teorema}{Theorem}[section]

\newtheorem{cor}{Corollary}[section]
\newtheorem{lemma}{Lemma}[section]
\newtheorem{prop}{Proposition}[section]
\newtheorem{quest}{Question}
\theoremstyle{remark}
\newtheorem{oss}{Remark}[section]

\newlanguage\fakelanguage
\newcommand\cyr{\fontencoding{OT2}\fontfamily{wncyr}\selectfont
   \language\fakelanguage}
\DeclareTextFontCommand{\textcyr}{\cyr}

\numberwithin{equation}{section}

\DeclareMathOperator{\End}{End}

\usepackage{calligra}

\makeatletter
\newsavebox{\@brx}
\newcommand{\llangle}[1][]{\savebox{\@brx}{\(\m@th{#1\langle}\)}%
  \mathopen{\copy\@brx\kern-0.5\wd\@brx\usebox{\@brx}}}
\newcommand{\rrangle}[1][]{\savebox{\@brx}{\(\m@th{#1\rangle}\)}%
  \mathclose{\copy\@brx\kern-0.5\wd\@brx\usebox{\@brx}}}
\makeatother

\title[Coalescence of $QH^\bullet(\mathbb G(\MakeLowercase{k,n}))$ and the Distribution of Prime Numbers]
{Coalescence Phenomenon of Quantum Cohomology of Grassmannians and the Distribution of Prime Numbers}
\author{Giordano Cotti$^{(\dag)}$}
\address[$\dag$]{SISSA, Via Bonomea, 265 - 34136 Trieste ITALY}
\email{gcotti@sissa.it}

\begin{document}

\begin{abstract}
The occurrence and frequency of a phenomenon of resonance (namely the coalescence of some Dubrovin canonical coordinates) in the locus of Small Quantum Cohomology of complex Grassmannians is studied. It is shown that surprisingly this frequency is strictly subordinate and highly influenced by the distribution of prime numbers. Two equivalent formulations of the Riemann Hypothesis are given in terms of numbers of complex Grassmannians without coalescence: the former as a constraint on the disposition of singularities of the analytic continuation of the Dirichlet series associated to the sequence counting non-coalescing Grassmannians, the latter as asymptotic estimate (whose error term cannot be improved) for their distribution function.
\end{abstract}

\maketitle
\tableofcontents

\newpage
\section*{Notations}
Given two natural numbers $0<k<n$ we will denote by $\mathbb G(k,n)$ the Grassmannian of $k$-dimensional $\mathbb C$-subspaces of $\mathbb C^n$. Thus $\mathbb G(1,n)=\mathbb P^{n-1}_{\mathbb C}$.\newline

In what follows we will use the following notations for number theoretical functions:
\begin{itemize}
\item $P_1(n):=\min\left\{p\in\mathbb N\colon p\text{ is prime and } p|n\right\},\ n\geq 2$;
\item for real positive $x,y$ we define
\[\Phi(x,y):=\operatorname{card}\left(\left\{n\leq x\colon n\geq2,\ P_1(n)> y\right\}\right);
\]
\item $\pi_\alpha(n):=\sum_{\substack{p\text{ prime}\\ p\leq n}}p^\alpha,\quad \alpha\geq 0;$
\item $\zeta(s)$ is the Riemann $\zeta$-function;
\item $\zeta(s,k)$ will denote the truncated Euler product
\[\zeta(s,k):=\prod_{{\substack{p\textnormal{\text{ prime}}\\ p\leq k}}}\left(1-\frac{1}{p^s}\right)^{-1},\quad k\in\mathbb R_{>0},\ s\in\mathbb C\setminus\left\{0\right\}.
\]
\item $\zeta_P(s)$ is the Riemann prime  $\zeta$-function, defined on the half-plane $\operatorname{Re}(s)>1$ by the series 
\[\zeta_P(s):=\sum_{p\textnormal{ prime}}\frac{1}{p^s};
\]
\item $\zeta_{P,k}(s)$ will denote the partial sums
\[\zeta_{P,k}(s):=\sum_{\substack{p\textnormal{ prime}\\ p\leq k}}\frac{1}{p^s};
\]
\item $\omega\colon \mathbb R_{\geq 1}\to\mathbb R$ is the Buchstab function (\cite{buchstab}), i.e. the unique continuous solution of the delay differential equation
\[\frac{d}{du}(u\omega(u))=\omega(u-1),\quad u\geq 2,
\]with the initial condition 
\[\omega(u)=\frac{1}{u},\quad\text{for }1\leq u\leq 2.
\] 
\end{itemize}$ $\newline

If $f,g\colon\mathbb R_+\to\mathbb R$, with $g$ definitely strictly positive, we will write 
\begin{itemize}
\item $f(x)=\Omega_+(g(x))$ to denote
\[\underset{x\to\infty}{\lim\sup}\ \frac{f(x)}{g(x)}>0;
\]
\item $f(x)=\Omega_-(g(x))$ to denote
\[\underset{x\to\infty}{\lim\inf}\ \frac{f(x)}{g(x)}<0;
\]
\item $f(x)=\Omega_\pm(g(x))$ if both $f(x)=\Omega_+(g(x))$ and $f(x)=\Omega_-(g(x))$ hold.
\end{itemize}

\newpage
\section{Introduction and Results}

In this paper we exhibit a direct connection between the theory of Frobenius Manifolds, more precisely the Gromov-Witten and Quantum Cohomology theories, and one of the most ancient problems lying at the heart of Mathematics, the distribution of prime numbers. We study a phenomenon of resonance in the small quantum cohomology of complex Grassmannians (consisting in the coalescence of some natural parameters there defined), and show that the occurrence and frequency of this phenomenon is surprisingly related to the distribution of prime numbers. This relation is so strict that it leads to (at least) two equivalent formulations of the famous Riemann Hypothesis: the former is given as a constraint on the disposition of the singularities of a generating function of the numbers of Grassmannians not presenting the resonance, the latter as an (essentially optimal) asymptotic estimate for a distribution function of the same kind of Grassmannians. Besides their geometrical-enumerative meaning, three point genus zero Gromov-Witten invariants of complex Grassmannians implicitly contain information about the distribution of prime numbers. This mysterious relation deserves further investigations.\newline      

Born in the last decades of the XX-th century, in the middle of the creative impetus for a mathematically rigorous foundations of Mirror Symmetry, the theory of Frobenius Manifolds (\cite{dubro1}, \cite{dubro0}, \cite{dubro2}, \cite{manin}, \cite{hertling}, \cite{sabbah}) seems to be characterized by a sort of \emph{universality} (see \cite{dubro3}): this theory, in some sense, is able to unify in a unique, rich, geometrical and analytical description many aspects and features shared by the theory of Integrable Systems, Singularity Theory, Gromov-Witten Invariants, the theory of Isomondromic Deformations and Riemann-Hilbert Problems, as well as the theory of special functions like Painlevé Transcendents.\newline

Originally introduced by physicists (\cite{vafa}), in the context of $N=2$ Supersymmetric Field Theories and mirror phenomena, the Quantum Cohomology of a complex projective variety $X$ (or more in general a symplectic manifold \cite{salamcduff}) is a family of deformations of its classical cohomological algebra structure defined on $H^\bullet(X):=\bigoplus_k H^{k}(X;\mathbb C)$, and parametrized over an open domain $\mathcal D\subseteq H^\bullet(X)$: the fiber over $p\in \mathcal D$ is identified with the tangent space $T_p\mathcal D\cong H^\bullet (X)$. This is exactly the prototype of a Frobenius manifolds, namely a manifold, endowed with a flat metric, on whose tangent spaces there is a well defined commutative, associative, unitary algebra structure, satisfying also some more compatibility conditions. The structure constants of the quantum deformed algebras are given by (third derivatives of) a generating function $F^X_0$ of Gromov-Witten Invariants of genus 0 of $X$: these rational numbers morally ``count'' (modulo parametrizations) algebraic/pseudo-holomorphic curves of genus 0 on $X$, with a fixed degree, and intersecting some fixed subvarieties of $X$.\newline

Focusing on the case of complex Grassmannians, $X=\mathbb G(k,n)$, in what follows we study the occurrence of a phenomenon of coalescence of some numerical parameters on their small quantum cohomology (i.e. $\mathcal D\cap H^2(X)$). This set of parameters defines a system of coordinates (called \emph{Dubrovin canonical coordinates}) near any point whose corresponding Frobenius algebra is \emph{semisimple}. It is important to keep in mind that these parameters have a double algebro-geometric meaning:
\begin{enumerate}
\item at each point $p\in\mathcal D$ their vector fields coincide with the idempotents of the algebra structure defined on $T_p\mathcal D$;
\item they are the eigenvalues\footnote{The canonical coordinates are not uniquely determined by the requirement (1) alone: there is a shift ambiguity. We fix this freedom by requirement (2).} of an operator of multiplication by a distinguished vector field (the \emph{Euler vector field}) which codifies the grading structure of the algebras. Notice that, along the small quantum locus, the Euler vector field coincides with the first Chern class $c_1(X)$. 
\end{enumerate}
Moreover, this coalescence of canonical coordinates can be interpreted as a ``delicate point''\footnote{In the theory of Painlevé equations they are called \emph{critical points}.} in the local description of semisimple Frobenius manifolds as spaces of deformation parameters for isomonodromic families of linear differential systems with rational coefficients in complex domains, at least as exposed in \cite{dubro0} and \cite{dubro2}. Such a description is carried on in all details in \cite{CG}, for the general analytic case, and in \cite{CDG} for the specific Frobenius case. For simplicity, we will call \emph{coalescing} a Grassmannian such that some of the Dubrovin canonical coordinates coalesce.
\newline

The questions, to which we answer in the present paper, are the following:

\begin{enumerate}
\item[(i)] For which $k,n$ the Grassmannian $\mathbb G(k,n)$ is coalescing?
\item[(ii)] How frequent is the phenomenon of coalescence among all Grassmannians?
\end{enumerate}$ $\newline

Let us summarize some of the main results obtained.\newline

\textbf{Theorem} (cf. Theorems \ref{main}, \ref{main0.5}, \ref{main1}, \ref{main2} and Corollaries \ref{cor1}, \ref{cor2} for more details)\vspace{5 mm}

\noindent
\textbf{PART I}
\emph{ The complex Grassmannian $\mathbb G(k,n)$ is coalescing if and only if $P_1(n)\leq k\leq n-P_1(n)$. In particular, all Grassmannians of proper subspaces of $\mathbb C^p$, with $p$ prime, are not coalescing.}\vspace{5 mm}

\noindent
\textbf{PART II}\emph{Let us denote by $\tilde{\textnormal{\textcyr{l}}}_n$, for $n\geq 2$, the number of non-coalescing Grassmannians of proper subspaces of $\mathbb C^n$, i.e.}
\[\tilde{\textnormal{\textcyr{l}}}_n:=\operatorname{card}\left\{k\colon \mathbb G(k,n)\text{ \emph{is not coalescing}}\right\},
\]\emph{and let} 
\[\widetilde{\textnormal{\textcyr{L}}}(s):=\sum_{n=2}^\infty\frac{\tilde{\textnormal{\textcyr{l}}}_n}{n^s}
\]\emph{be the associated Dirichlet series. $\widetilde{\textnormal{\textcyr{L}}}(s)$ is absolutely convergent in the half-plane $\operatorname{Re}(s)> 2$, where it can be represented by the infinite series involving the Riemann zeta function and the truncated Euler products
\[\widetilde{ \textnormal{\textcyr{L}}}(s)=\sum_{p\textnormal{ prime}}\frac{p-1}{p^s}\left(\frac{2\zeta(s)}{\zeta(s,p-1)}-1\right).
\]
By analytic continuation, $\widetilde{\textnormal{\textcyr{L}}}(s)$ can be extended to (the universal cover of) the punctured half-plane $$\left\{s\in\mathbb C\colon \operatorname{Re}(s)>\overline{\sigma}\right\}\setminus\left\{s=\frac{\rho}{k}+1\colon\begin{aligned}
&\rho \text{ pole or zero of }\zeta(s),\\
k&\text{ squarefree positive integer}
\end{aligned}\right\},$$ $$\overline{\sigma}:=\underset{n\to\infty}{\lim\sup}\ \frac{1}{\log n}\cdot \log\left(\sum_{\substack{k\leq n \\k\textnormal{ composite}}}\tilde{\textnormal{\textcyr{l}}}_k\right),\quad 1\leq\overline\sigma\leq\frac{3}{2},$$ having logarithmic singularities at the punctures. }\newline

\noindent
\emph{In particular, we have the equivalence of the following statements:
\begin{itemize}
\item \textnormal{(RH)} all non-trivial zeros of the Riemann zeta function $\zeta(s)$ satisfy $\operatorname{Re}(s)=\frac{1}{2}$;
\item the derivative $\widetilde{\textnormal{\textcyr{L}}}'(s)$ extends, by analytic continuation,  to a meromorphic function in the half-plane $\frac{3}{2}<\operatorname{Re}(s)$ with a single pole of oder one at $s=2$.
\end{itemize}} $ $

\noindent
\emph{At the point $s=2$ the following asymptotic estimate holds
\[\widetilde{\textnormal{\textcyr{L}}}(s)=\log\left(\frac{1}{s-2}\right)+O(1),\quad s\to 2,\quad \operatorname{Re}(s)>2.
\]As a consequence, we have that
\[\sum_{k=2}^n\tilde{\textnormal{\textcyr{l}}}_k\sim\frac{1}{2}\frac{n^2}{\log n},
\]which means that non-coalescing Grassmannians are \emph{rare}.
}\vspace{5 mm}
\begin{figure}[ht]
        \centering
        \includegraphics[scale=.55]{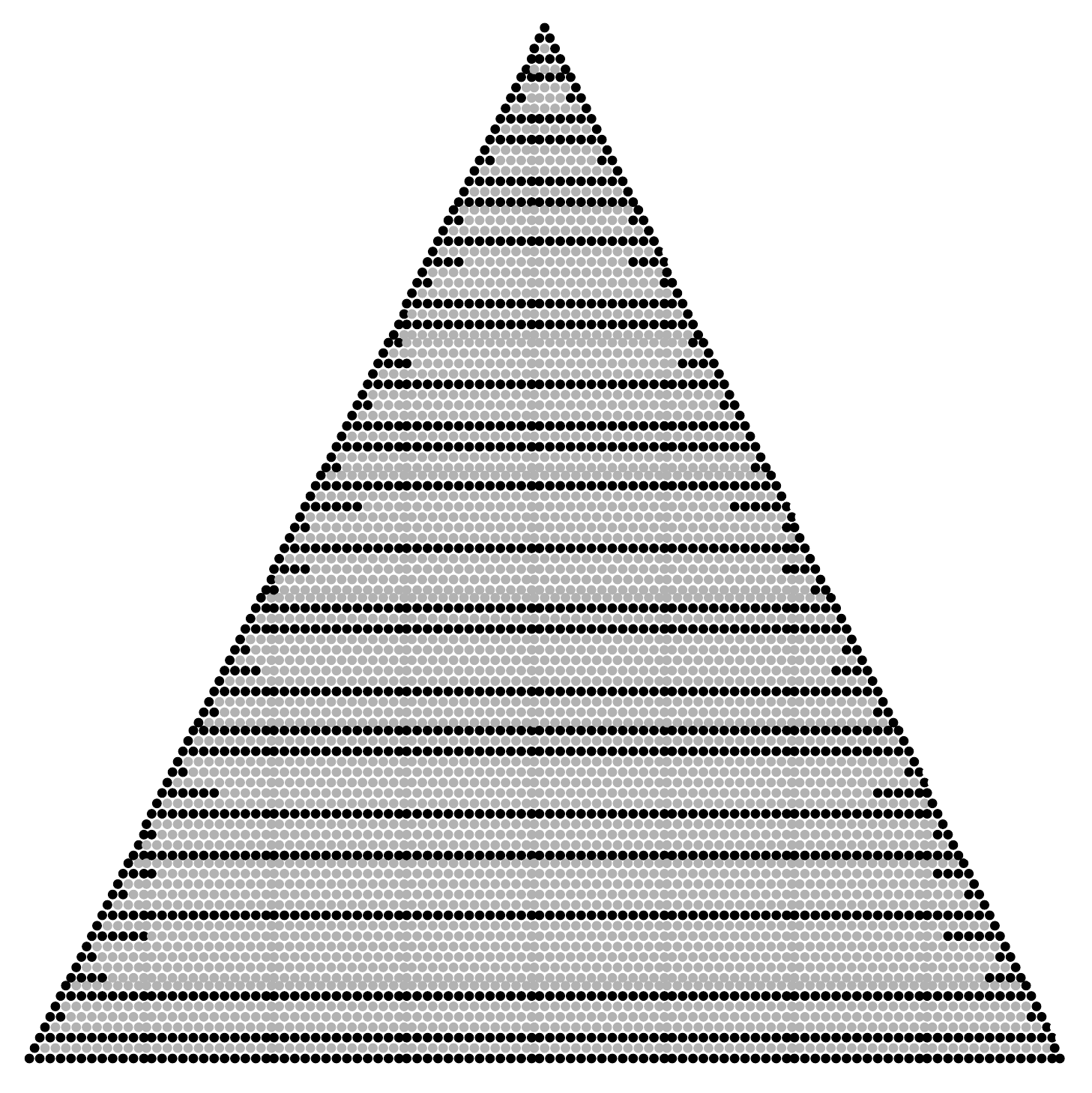}
        \caption{In this figure we represent complex Grassmannians as disposed in a Tartaglia-Pascal triangle: the $k$-th element (from the left) in the $n$-th row (from the top of the triangle) represents the Grassmannian $\mathbb G(k,n+1)$, where $n\leq 102$. The dots colored in black represent \emph{non-coalescing} Grassmannians, while the dots colored in gray the \emph{coalescing} ones. The reader can note that black dots are \emph{rare} w.r.t. the gray ones, and that the black lines correspond to Grassmannians of subspaces in $\mathbb C^p$ with $p$ prime.}
   \end{figure}

\noindent
\textbf{PART III}\emph{ For $x\in\mathbb R_{\geq 4}$ we define
\[\widehat{\mathscr H}(x):=\operatorname{card}\left\{n\colon\begin{aligned}
&\quad 2\leq n\leq x\text{ is such that }\mathbb G(k,n)\\
&\text{ is not coalescing for }1\leq k\leq [x^\frac{1}{2}]+1
\end{aligned}\right\}.
\]}\emph{We have the following results:}
\begin{enumerate}
\item[(a)] \emph{for any $\kappa>1$, the following integral representations\footnote{The integral must be interpreted as a Cauchy Principal Value.} hold 
\[\widehat{\mathscr H}(x)=\frac{1}{2\pi i}\int_{\Lambda_\kappa}\left[\left(\frac{\zeta(s)}{\zeta\left(s,x^\frac{1}{2}+1\right)}-1\right)-\zeta_{P,2x^\frac{1}{2}+1}(s)+\zeta_{P,x^\frac{1}{2}+1}(s)\right]\frac{x^s}{s}ds, 
\]
\[\widehat{\mathscr H}(x)=\frac{1}{2\pi i}\int_{\Lambda_\kappa}\left[\left(\frac{\zeta(s)}{\zeta\left(s,x^\frac{1}{2}+1\right)}-1\right)x^s+\zeta_P(s)\left((x^\frac{1}{2}+1)^s-(2x^\frac{1}{2}+1)^s\right)\right]\frac{ds}{s},
\]
both valid for $x\in\mathbb R_{\geq 2}\setminus\mathbb N,$ and where $\Lambda_\kappa:=\left\{\kappa+it\colon t\in\mathbb R\right\}$ is the line oriented from $t=-\infty$ to $t=+\infty$.}

\item[(b)] \emph{The function $\widehat{\mathscr H}$ admits the following asymptotic estimate:
\[\widehat{\mathscr H}(x)=\int_{0}^x\frac{dt}{\log t}+O\left(x^\Theta\log x\right),\quad \text{where }\Theta:=\sup\left\{\operatorname{Re}(\rho)\colon \zeta(\rho)=0\right\}.
\]}
\end{enumerate}
\emph{
Hence, it is clear the equivalence of \textnormal{(RH)} with the (essentially optimal) estimate with $\Theta=\frac{1}{2}$. 
}\vspace{7 mm}

Question (i) has already been addressed in \cite{gamma1} (Remark 6.2.9): it is claimed, but not proved, that the condition $\gcd(\min(k,n-k)!,n)>1$ (which is equivalent to the condition $P_1(n)\leq k\leq n-P_1(n)$) is a necessary condition for coalescence of some canonical coordinates in the small quantum locus of $\mathbb G(k,n)$. 

\subsection{Plan of the paper.} In Section 2 we summarize some basic notions about Frobenius Manifolds, Gromov-Witten Theory and Quantum Cohomology. In Section 3 we expose a description, valid both in the classical and in the quantum setup, of the cohomology of Grassmannians as alternate products of cohomology of Projective Spaces.\newline In Section 4 the coalescence phenomenon is completely characterized, and a generating function (Dirichlet series) for the numbers of non-coalescing Grassmannians is introduced: from the study of the behavior of this function near its singularities, we deduce the \emph{rareness} (i.e. zero density) of the non-coalescence.\newline In Section 5 we introduce and study some distribution functions for non-coalescing Grassmannians: some integral representations and asymptotic expansions are found. It is shown that RH can be reformulated as an (essentially optimal) estimate for one of the distribution function introduced. \vspace{5 mm}

\noindent
\textbf{Acknowledgements.}  The author is grateful to Boris Dubrovin and Davide Guzzetti for useful discussions and encouragement, to Alberto Perelli and Gérald Tenenbaum for the friendly e-mail conversations, to Don Zagier for critical comments, and to Michael Zieve for pointing out a Theorem due to H.B. Mann (Theorem \ref{mann}), used in the proof of Proposition \ref{propchiave}.

\section{Frobenius Manifolds and Quantum Cohomology}
Introduced and extensively developed by B. Dubrovin in \cite{dubro0}, \cite{dubro1}, \cite{dubro2} in order to give a differential geometrical description of the WDVV-system of equations obtained in two dimensional topological field theories (\cite{DVV}, \cite{witten}), Frobenius Manifolds are, roughly speaking, smooth/analytic manifolds whose tangent spaces are endowed with an associative, commutative and unitary algebra\footnote{In the smooth and real analytic cases, the ground field of the algebra is $\mathbb R$, in the complex analytic case (on which we will focus in the present paper) is $\mathbb C$.} structure, smoothly/analytically depending on the point. Furthermore, in order to be \emph{Frobenius}, these algebras structures must satisfy a compatibility condition with respect to a symmetric bilinear non-degenerate form, simply called \emph{metric}\footnote{If $M$ is a complex analytic Frobenius manifold, the metric is defined on the holomorphic tangent bundle $TM$, and it is just $\mathcal O_M$-bilinear.}: if $M$ is a Frobenius manifold, with metric $\eta$, and we denote by $\circ_p$ the product defined on the tangent space $T_pM$ at the point $p\in M$, then the required compatibility condition is
\begin{equation}\label{04-08-16}\eta(a\circ_pb,c)=\eta(a,b\circ_pc)\quad\text{for all }a,b,c\in T_pM,\ p\in M.
\end{equation}
In the remaining part of the paper, we will consider only complex analytic Frobenius manifolds. It could be useful for the reader to keep in mind the following elementary example of Frobenius algebra.

\begin{es}\label{es1}
Let $X$ be a smooth compact manifold of even dimension, with vanishing odd cohomology, i.e. $H^{2k+1}(X)\cong 0$ for all $k\geq 0$. Let us consider its classical cohomology ring $A:=\bigoplus H^k(X;\mathbb C)$, endowed with the $\cup$-product, and the Poincaré metric $\eta$ defined by
\[\eta(\alpha,\beta):=\int_X\alpha\cup \beta.
\]Then, the resulting algebra structure is associative, commutative, unitary, there is a well defined $\mathbb C$-bilinear symmetric form which is non-degenerate (Poincaré Duality Theorem), with respect to which the compatibility condition \eqref{04-08-16} is trivially satisfied. $A$ is the classical Frobenius cohomological algebra associated to the manifold $X$. From a physical point of view, this algebra describes the matter sector of the topological $\sigma$-model with target space $X$. \end{es}

One of the main features of the algebra $A$ is that it is a naturally \emph{Frobenius graded algebra}: there exists a linear morphism $Q\colon A\to A$, called \emph{grading operator}, such that
\[Q(\alpha\cup\beta)=Q(\alpha)\cup\beta+\alpha\cup Q(\beta),
\]
\[\eta(Q(\alpha),\beta)+\eta(\alpha,Q(\beta))=d\cdot\eta(\alpha,\beta),
\]where $d$ is a number called the \emph{charge} of the algebra. In Example \ref{es1}, the operator $Q$ is defined as the map acting on a homogeneous basis $(T_i)_i$ of $A$ as $Q(T_i)=\frac{1}{2}\deg(T_i)\cdot T_i$, and the charge is $d=\frac{1}{2}\dim_{\mathbb R}X$.

If we restrict to the case of Frobenius algebras with \emph{diagonalizable} grading operators, by imposing a flatness condition for the Levi-Civita connection associated to $\eta$, we can extend the grading structure on a whole Frobenius manifold by the choice of a distinguished affine vector field, the so-called \emph{Euler vector field.} We give now the complete and detailed definition.
\begin{defi}A \emph{Frobenius manifold} structure on a complex manifold $M$ of dimension $n$ is defined by giving
\begin{enumerate}
\item[(FM1)] a symmetric $\mathcal O(M)$-bilinear metric tensor $\eta\in\Gamma\left(\bigodot^2T^*M\right)$, whose corresponding Levi-Civita connection $\nabla$ is flat;
\item[(FM2)] a $(1,2)$-tensor $c\in\Gamma\left(TM\otimes\bigodot^2T^*M\right)$ such that
\begin{itemize}
\item $c^\flat\in\Gamma\left(\bigodot^3T^*M\right)$,
\item $\nabla c^\flat\in\Gamma\left(\bigodot^4T^*M\right)$;
\end{itemize}
\item[(FM3)] a vector field $e\in\Gamma(TM)$, called the \emph{unity vector field}, such that
\begin{itemize}
\item the bundle morphism $c(-,e,-)\colon TM\to TM$ is the identity morphism,
\item $\nabla e=0$;
\end{itemize}
\item[(FM4)] a vector field $E\in\Gamma(TM)$, called the \emph{Euler vector field}, such that
\begin{itemize}
\item $\frak L_Ec=c$,
\item $\frak L_E\eta=(2-d)\cdot \eta$, where $d\in\mathbb C$ is called the \emph{charge} of the Frobenius manifold.
\end{itemize}
\end{enumerate}
\end{defi}

The tensor $c$ is the tensor of constants structure of a Frobenius algebra on each tangent space: the multiplication of vector fields will be denoted by $\circ$. We leave as an easy exercise for the reader to deduce from the axioms above that the resulting algebra on each tangent space is Frobenius. Being the connection $\nabla$ flat, there exist local flat coordinates, that we denote $(t^\alpha)_{\alpha}$, w.r.t. which the metric $\eta$ is constant. Moreover, observe that because of flatness and the conformal Killing condition, the Euler vector field is affine, i.e. $\nabla\nabla E=0$. Without loss of generality, we can mark the first flat coordinate $t^1$ in such a way that 
\begin{itemize}
\item $\frac{\partial}{\partial t^1}$ coincides with the unity vector field, 
\item and that the local expression of the Euler vector field is $E=\sum_\nu\left[(1-q_\nu)t^\nu+r^\nu\right]\partial_\nu$, with $q_\nu,r^\nu\in\mathbb C$, $q_1=0$  and $r^\nu\neq 0$ if and only if $q_\nu=1$.
\end{itemize}
Since the metric $\eta$ is flat, in flat local coordinates $(t^\alpha)_{\alpha}$ the connection $\nabla$ coincides with partial derivatives: so, the condition $\nabla c^\flat\in\Gamma\left(\bigodot^4T^*M\right)$ means that $\partial_{\alpha}c_{\beta\gamma\delta}$ is symmetric in all indices. This implies the local existence of a function $F$ such that
\[c_{\alpha\beta\gamma}=\partial_\alpha\partial_\beta\partial_\gamma F.
\]
The associativity of the algebra, together with its graded structure, are equivalent to the following conditions for $F$, called WDVV-equations:
\begin{equation}\label{WDVV1}
F_{\alpha\beta\gamma}\eta^{\gamma\delta}F_{\delta\epsilon\nu}=F_{\nu\beta\gamma}\eta^{\gamma\delta}F_{\delta\epsilon\alpha}, \quad F_{\alpha\beta\gamma}:=\partial_\alpha\partial_\beta\partial_\gamma F.
\end{equation}
\begin{equation}\label{WDVV2}\eta_{\alpha\beta}=F_{1\alpha\beta},\quad \frak L_EF=(3-d)\cdot F+Q(t),
\end{equation}where $Q(t)$ denotes a quadratic polynomial in $t^\alpha$'s. Giving a solution of the WDVV problem \eqref{WDVV1}, \eqref{WDVV2} is equivalent to locally defining a Frobenius manifold structure (for more details see \cite{dubro1}, \cite{dubro2}).

\subsection{Semisimple Frobenius Manifolds} 
We will focus now on a particularly interesting and vast class of Frobenius manifolds, the class of \emph{semisimple} Frobenius manifolds.
\begin{defi}
A point $p$ of a Frobenius manifold $M$, of complex dimension $n$, is said to be a \emph{semisimple point} if the corresponding Frobenius algebra $(T_pM,\eta|_p,\circ_p)$ satisfies one of the following equivalent conditions:
\begin{enumerate}
\item it is without nilpotents;
\item it is isomorphic to $\mathbb C^n$;
\item it admits a basis of orthogonal idempotent vectors $\varphi_1,\dots, \varphi_n$: \[\varphi_i\circ_p\varphi_j=\delta_{ij}\varphi_i,\quad \eta_p(\varphi_i,\varphi_j)=\eta_p(\varphi_i,\varphi_i)\delta_{ij};\]
\item it admits a \emph{regular} vector, i.e. a vector $v$ such that the operator $v\circ_p (-)\colon T_pM\to T_pM$ has simple spectrum.
\end{enumerate} A Frobenius manifold with an open dense subset of semisimple points is called \emph{semisimple}\footnote{A sufficient condition for a complex analytic Frobenius manifold to be semisimple is to have \emph{at lest} one semisimple point: it is clear, indeed, that the semisimplicity condition is open. For a proof of the density of the open set of semisimple points see \cite{hertling}.}.
\end{defi}

Locally, near semisimple points, a distinguished system of (non-flat) coordinates is always well-defined:
\begin{teorema}[B. Dubrovin \cite{dubro2}; see also \cite{CDG}]Let $M$ be a Frobenius manifold, and $p_0\in M$ a semisimple point. On any sufficiently small, simply-connected, open neighborhood $U$ of $p_0$, made of semisimple points, a coherent labeling $(u_1,\dots, u_n)$ of the eigenvalues of the operators $\mathcal U:=E\circ_p (-)\colon T_pM\to T_pM$, $p\in U$, can be used as holomorphic local coordinates such that the coordinate vector fields $$\frac{\partial}{\partial u_1},\dots,\frac{\partial}{\partial u_n}$$ are the orthogonal idempotent vector fields at any point $p\in U$.
\end{teorema}

We will refer to this particular system of coordinates as the \emph{system of Dubrovin canonical coordinates}. One of the main features of the theory of semisimple Frobenius manifolds is that their whole structure can be locally parametrized by a finite number of parameters, or \emph{local moduli}. A particularly effective method of local parametrization of semisimple Frobenius structures consists in their (local) identification with spaces of independent deformation parameters for isomonodromic families of equations in complex domains with rational coefficients. If $M$ is a Frobenius manifold, and $p_0\in M$ is a point with \emph{pairwise distinct} canonical coordinates (i.e. $u_i(p_0)\neq u_j(p_0)$ for $i\neq j$; this implies that $p_0$ is semisimple), then, for a sufficiently small neighborhood $\mathcal N$ of $p_0$, we can associate to any point $p\in\mathcal N$ a differential system on $\mathbb P^1(\mathbb C)\setminus\left\{0,\infty\right\}$ of the form
\begin{equation}\label{iso1}\frac{dY}{dz}=\left(U(p)+\frac{V(p)}{z}\right)Y,\quad Y(z)\in M_n(\mathbb C),\ z\in\widehat{\mathbb C\setminus{0}},
\end{equation}such that
\begin{itemize}
\item $U(p):=\operatorname{diag}(u_1(p),\dots, u_n(p))$,
\item the matrix $V(p)$ is anti-symmetric,
\item the monodromy data of solutions of the equation do not depend on $p$.
\end{itemize}
We refer the reader to \cite{dubro1},\cite{dubro2} and \cite{guzzetti2} for more details, and explicit formulae for the local reconstruction of the Frobenius structure starting from the monodromy data of the isomonodromic family \eqref{iso1}. 

The extension of this local description of Frobenius manifolds, as spaces of deformation parameters of isomonodromic families of equations, near semisimple points with \emph{coalescing} canonical coordinates, is quite delicate, and it is treated in all details in \cite{CG} (for the general analytic theory) and in \cite{CDG} (for the specific case Frobenius). 

\subsection{Gromov-Witten Theory and Quantum Cohomology}
Let $X$ be a smooth projective complex variety. In order not to deal with superstructures, we will suppose for simplicity that the variety $X$ has vanishing odd cohomology, i.e. $H^{2k+1}(X;\mathbb C)\cong 0$ for $0\leq k$. Let us fix a homogeneous basis $(T_0,T_1,\dots, T_N)$ of $H^\bullet(X;\mathbb C)=\bigoplus_{k}H^{2k}(X;\mathbb C)$ such that
\begin{itemize}
\item $T_0=1$ is the unity of the cohomology ring;
\item $T_1,\dots, T_r$ span $H^2(X;\mathbb C)$.
\end{itemize}
We will denote by $\eta\colon H^\bullet(X;\mathbb C)\times H^\bullet(X;\mathbb C)\to\mathbb C$ the Poincaré metric
\[\eta(\xi,\zeta):=\int_X\xi\cup\zeta,
\]and in particular 
\[\eta_{\alpha\beta}:=\int_XT_\alpha\cup T_\beta.
\]

If $\beta\in H_2(X;\mathbb Z)/\text{torsion}$, we denote by $\overline{\mathcal M}_{g,n}(X,\beta)$ the Kontsevich-Manin moduli stack of $n$-pointed, genus $g$ stable maps to $X$ of degree $\beta$, which parametrizes:
\begin{itemize}
\item equivalence classes of pairs $((C_g,\bold x); f)$, where $(C_g,\bold x)$ is an $n$-pointed algebraic curve of genus $g$, with at most nodal singularities and with $n$ marked points $\bold x=(x_1,\dots, x_n)$, and $f\colon C_g\to X$ is a morphism such that $f_*[C_g]\equiv \beta$. Two pairs $((C_g,\bold x);f)$ and $((C'_g,\bold {x'});f')$ are defined to be equivalent if there exists a bianalytic map $\varphi\colon C_g\to C'_g$ such that $\varphi(x_i)=x'_i$, for all $i=1,\dots, n$, and $f'=\varphi\circ f$.

\item The morphisms $f$ are required to be \emph{stable}: if $f$ is constant on any irreducible component of $C_g$, then that component should have only a finite number of automorphisms as pointed curves (in other words, it must have at least 3 distinguished points, i.e. points that are either nodes or marked ones). 
\end{itemize}

We will denote by $\operatorname{ev_i}\colon \overline{\mathcal M}_{g,n}(X,\beta)\to X\colon ((C_g,\bold x); f)\mapsto f(x_i)$ the naturally defined evaluations maps, and by $\psi_i\in H^2(\overline{\mathcal M}_{g,n}(X,\beta);\mathbb Q)$ the Chern classes of tautological cotangent line bundles $$\mathcal L_i\to \overline{\mathcal M}_{g,n}(X,\beta),\quad \mathcal L_i|_{((C_g,\bold x); f)}=T^*_{x_i}C_g,\quad \psi_i:=c_1(\mathcal L_i).$$ 
Using the construction of \cite{fantechi} of a \emph{virtual fundamental class} $[\overline{\mathcal M}_{g,n}(X,\beta)]^{\text{virt}}$ in the Chow ring $A_*(\overline{\mathcal M}_{g,n}(X,\beta))$, and of degree equal to the expected dimension
\[[\overline{\mathcal M}_{g,n}(X,\beta)]^{\text{virt}}\in A_D(\overline{\mathcal M}_{g,n}(X,\beta)), \quad D=(1-g)(\dim_{\mathbb C}X-3)+n+\int_Xc_1(X),
\]a good theory of intersection is allowed on the Kontsevich-Manin moduli stack. 

We can thus define the \emph{Gromov-Witten invariants (with descendants) of genus $g$, with $n$ marked points and of degree $\beta$} of $X$ as the integrals (whose values are rational numbers)
\begin{equation}\label{gw1}\langle\tau_{d_1}\gamma_1,\dots,\tau_{d_n}\gamma_n\rangle_{g,n,\beta}^X:=\int_{[\overline{\mathcal M}_{g,n}(X,\beta)]^{\text{virt}}}\prod_{i=1}^n\operatorname{ev}_i^*(\gamma_i)\cup\psi_i^{d_i},
\end{equation}
\[ \gamma_i\in H^\bullet(X;\mathbb C),\quad d_i\in\mathbb N,\quad i=1,\dots, n.
\]Since by \emph{effectiveness} (see \cite{manin}, \cite{kon}) the integral is non-vanishing only for effective classes $\beta\in\operatorname{Eff}(X)\subseteq H_2(X;\mathbb Z)$, the generating function of rational numbers \eqref{gw1}, called \emph{total descendent potential} (or also \emph{gravitational Gromov-Witten potential}, or even \emph{Free Energy}) of \emph{genus $g$} is defined as the formal series
\[\mathcal F^X_g(\gamma,\bold Q):=\sum_{n=0}^\infty\sum_{\beta\in \text{Eff}(X)}\frac{\bold Q^\beta}{n!}\langle\underbrace{\gamma.\dots,\gamma}_{n\text{ times}}\rangle_{g,n,\beta}^X,
\]
where we have introduced (infinitely many) coordinates $\bold{t}:=(t^{\alpha,p})_{\alpha,p}$
\[\gamma=\sum_{\alpha,p}t^{\alpha,p}\tau_pT_{\alpha},\quad \alpha=0,\dots, N,\ p\in\mathbb N,
\] and formal parameters
\[\bold Q^\beta:=Q_1^{\int_\beta T_1}\cdot\dots\cdot Q_r^{\int_\beta T_r},\quad Q_i \text{'s elements of the Novikov ring }\Lambda:=\mathbb C[\![Q_1,\dots,Q_r]\!].
\]
The free energy $\mathcal F^X_g\in\Lambda[\![{\bold t}]\!]$ can be seen a function on the \emph{large phase-space}, and restricting the free energy to the \emph{small phase space} (naturally identified with $H^\bullet(X;\mathbb C)$), 
\[F^X_g(t^{1,0},\dots, t^{N,0}):=\mathcal F^X_g(\bold t)|_{t^{\alpha,p}=0,\ p>0},
\]
one obtains the generating function of the Gromov-Witten invariants of genus $g$.

By the Divisor axiom, the genus $0$ Gromov-Witten potential $F^X_0(t)$, can be seen as an element of the ring $\mathbb C[\![t^0,Q_1e^{t^1},\dots,Q_re^{t^r},t^{r+1},\dots, t^N]\!]$: in what follows we will be interested in cases in which $F^X_0$ is the analytic expansion of an analytic function, i.e.
\[F^X_0\in\mathbb C\left\{t^0,Q_1e^{t^1},\dots,Q_re^{t^r},t^{r+1},\dots, t^N\right\}.
\]Without loss of generality, we can put $Q_1=Q_2=\dots=Q_r=1$, and $F^X_0(t)$ defines an analytic function in an open neighborhood $\mathcal D\subseteq H^\bullet(X;\mathbb C)$ of the point
\begin{align}\label{classical1}
t^i&=0,\quad i=0,r+1,\dots, N,\\
\label{classical2}
\operatorname{Re} t^i&\to -\infty,\quad i=1,2,\dots, r.
\end{align}
The function $F^X_0$ is a solution of WDVV equations (for a proof see \cite{kon}, \cite{manin}, \cite{cox}), and thus it defines an analytic Frobenius manifold structure on $\mathcal D$.

\begin{defi}
The Frobenius manifold structure defined on the domain of convergence $\mathcal D$ of the Gromov-Witten potential $ F^X_0$, solution of the WDVV problem, is called \emph{Quantum Cohomology} of $X$, and denoted by $QH^\bullet(X)$. Note that
\begin{itemize}
\item the flat metric is given by the Poincaré metric $\eta$;
\item the unity vector field is $T_0=1$, using the canonical identifications of tangent spaces $$T_p\mathcal D\cong H^\bullet(X;\mathbb C)\colon \partial_{t^\alpha}\mapsto T_\alpha;$$
\item the Euler vector field is
\[E:=c_1(X)+\sum_{\alpha=0}^N\left(1-\frac{1}{2}\deg T_\alpha\right)t^\alpha T_\alpha.
\]
\end{itemize}
By the expression \emph{small quantum cohomology} (or \emph{small quantum locus}) we denote the Frobenius structure attached to points in $\mathcal D\cap H^2(X;\mathbb C)$. 
\end{defi}

\begin{oss}
We have no general results guaranteeing the converge of the Gromov-Witten potential for a generic smooth projective variety $X$; however, for some classes of varieties, it is known that the sum defining $F^X_0$ at points with coordinates $t^0=t^{r+1}=\dots=t^N=0$ is finite. This is the case for
\begin{itemize}
\item Fano varieties,
\item varieties admitting a transitive action of a semisimple Lie group. 
\end{itemize}
For a proof see \cite{cox}. Notice that for these varieties the small quantum locus coincide with the whole space $H^2(X;\mathbb C)$. Conjecturally, for Calabi-Yau manifolds the series defining $F^X_0$ is convergent in a neighborhood of the classical limit point (see \cite{cox}, \cite{kon}). In case of convergence, the potential (and hence the whole Frobenius structure) can be maximally analytically continued to an unramified covering of a subdomain of $H^\bullet(X;\mathbb C)$. We refer to this global Frobenius structure as the \emph{Big Quantum Cohomology}.
\end{oss}

\begin{oss} At the classical limit point \eqref{classical1}, \eqref{classical2}, the algebra structure on the tangent spaces coincide with the classical cohomological algebra structure introduced in Example \ref{es1}. Indeed, the following is the structure of the potential:
\begin{enumerate}
\item by Point Mapping Axiom, the Gromov-Witten potential can be decomposed into a \emph{classical} term and a \emph{quantum} correction as follows
\begin{align*}F^X_0(\gamma)&=F_{\text{classical}}+F_{\text{quantum}}\\
&=\frac{1}{6}\int_X\gamma^3+\sum_{k=0}^\infty\sum_{\beta\in\operatorname{Eff}(X)\setminus\left\{0\right\}}\frac{1}{k!}\langle\underbrace{\gamma,\dots,\gamma}_{k\text{ times}}\rangle^X_{0,k,\beta},\quad\text{where }\gamma=\sum_{\alpha=0}^Nt^\alpha T_\alpha;
\end{align*}
\item the variable $t^0$ appears only in the classical term of $F^X_0$;
\item because of the Divisor axiom, the variables corresponding to cohomology degree 2 (i.e. $t^1,\dots, t^r$) appear in the exponential form in the quantum term; the Frobenius structure is $2\pi i$-periodic in the 2-nd cohomology directions: the structure can be considered as defined on an open region of the quotient $H^\bullet(X;\mathbb C)/2\pi i H^2(X;\mathbb Z)$. 
\end{enumerate} 
\end{oss}

\begin{oss}\label{ossemi}
The Frobenius algebra of Example \ref{es1} is not semisimple (it clearly contains nilpotents). By quantum deformation of the $\cup$-product, it may happen that the quantum cohomology is semisimple. There are no general results characterizing smooth projective varieties with semisimple quantum cohomology: however, for some classes of varieties such as
\begin{itemize}
\item some Fano threefolds \cite{ciolli},
\item toric varieties \cite{iri7},
\item some homogeneous spaces \cite{cmp10},
\end{itemize}
it has been proved that the small quantum locus is made of semisimple points. Grassmannians are among these varieties. More general homogeneous spaces may have non-semisimple small quantum cohomology (\cite{cmp10}, \cite{chaput-perrin}). Some necessary conditions for semisimplicity are given in \cite{hmt}; some sufficient conditions for other Fano varieties are given in \cite{perrin}. There is an intriguing conjecture (\cite{dubro0}, \cite{dubro4}) which relates the enumerative geometry of a projective variety $X$ with semisimple quantum cohomology with its derived category of coherent sheaves $\mathcal D^b(X)$: some more details and new results can be found in \cite{CDG} and \cite{CDG1}.
\end{oss}

\section{Quantum Satake Principle}
The quantum cohomology of Grassmannians has been one of the first cases that both physicists \cite{wit1} and mathematicians (see e.g. \cite{ber1}, \cite{bert}, \cite{buc1}) studied in details. In this section we expose an identification, valid both in the classical (\cite{martin}) and in the quantum setup (\cite{golymaniv}, \cite{BCFK}, \cite{gamma1}), of the cohomology of Grassmannians with an alternate product of the cohomology of Projective Spaces. This identification has been well known to physicists for long time: e.g. the reader can find an analogue description of the supersymmetric $\sigma$-model of $\mathbb G(k,n)$ in Section 8.3 and Appendix A of the paper \cite{cecotti-vafa}, on the classification of $N=2$ Supersymmetric Field Theories. In the context of the theory of Frobenius manifolds, such an identification has been generalized and axiomatized in \cite{KS} in the notion of \emph{alternate product of Frobenius manifolds}.
\subsection{Results on classical cohomology of Grassmannians} A classical reference for cohomology of Grassmannians is \cite{griffith}.
Let us introduce the following notations, used only in this section, to denote the (products of) complex flag manifolds
\[\mathbb P:=\mathbb P^{n-1}_{\mathbb C},\quad \Pi:=\underbrace{\mathbb P\times\dots\times\mathbb P}_{k\text{ times}}
\]
\[\mathbb G:=\mathbb G(k,n),\quad \mathbb F:=\operatorname{Fl}(1,2,\dots,k,n),
\] 
with $n\geq 2$, and $1\leq k< n$. If we fix an hermitian structure on $\mathbb C^n$ (e.g. the standard one), we have the following diagram:
\[\xymatrix{
&\mathbb F\ \ar[ld]_{p}\ar@{^{(}->}[rd]^{\iota}&\\
\mathbb G&&\Pi
}
\]where $p$ is the canonical projection, and $\iota$ the inclusion. Moreover, it is also defined a natural rational map «\emph{taking the span}»
\[\xymatrix{\Pi\ar@{-->}[r]&\mathbb G\colon\ (\ell_1,\dots,\ell_k)\mapsto \operatorname{span}\langle\ell_1,\dots,\ell_k\rangle},
\]whose domain is the image of $\iota$. On the manifold $\Pi$ we have $k$ canonical line bundles, denoted $\frak L_j$ for $j=1,\dots,k$, defined as the pull-back of the bundle $\mathcal O(1)$ on the $j$-th factor $\mathbb P$. If we denote $\frak V_1\subseteq\frak V_2\subseteq\dots\subseteq \frak V_k$ the tautological bundles over $\mathbb F$, we have that
\[\iota^*\frak L_j\cong (\frak V_j/\frak V_{j-1})^\vee.
\]
Denoting with the same symbol $x_i$ the Chern class $c_1(\frak L_i)$ on $\Pi$ and its pull-back $c_1(\iota^*\frak L_i)=\iota^*c_1(\frak L_i)$ on $\mathbb F$, we have
\[H^\bullet(\Pi;\mathbb C)\cong H^\bullet(\mathbb P;\mathbb C)^{\otimes k}\cong \frac{\mathbb C[x_1,\dots,x_k]}{\langle x_1^n,\dots x_k^n\rangle}\quad\text{(by Künneth Theorem),}
\]
\[H^\bullet(\mathbb F;\mathbb C)\cong\frac{\mathbb C[x_1,\dots,x_k]}{\langle h_{n-k+1},\dots, h_n\rangle},
\]where $h_j$ stands for the $j$-th complete symmetric polynomial in $x_1,\dots, x_k$. Since the classes $x_1,\dots, x_k$ are the Chern roots of the dual of the tautological bundle $\frak V_k$, we also have
\[H^{\bullet}\left(\mathbb G;\mathbb C\right)\cong\frac{\mathbb C[e_1,\dots, e_k]}{\langle h_{n-k+1},\dots,h_n\rangle}\cong \frac{\mathbb C[x_1,\dots,x_k]^{\frak S_k}}{\langle h_{n-k+1},\dots,h_n\rangle},
\]where the $e_j$'s are the elementary symmetric polynomials in $x_1,\dots, x_k$. This is the classical representation of the cohomology ring of the Grassmannian $\mathbb G$ with generators the Chern classes of the dual of the tautological vector bundle $\mathcal S$, and relations generated by the Segre classes of $\mathcal S$.\\
From this ring representation, it is clear that any cohomology class of $\mathbb G$ can be \emph{lifted} to a cohomology class of $\Pi$: we will say that $\tilde \gamma\in H^\bullet(\Pi;\mathbb C)$ is the lift of $\gamma\in H^\bullet(\mathbb G;\mathbb C)$ if $p^*\gamma=\iota^*\tilde\gamma$. The following integration formula allow us to express the cohomology pairings on $H^\bullet(\mathbb G;\mathbb C)$ in terms of the cohomology pairings on $H^\bullet(\Pi;\mathbb C)$.

\begin{teorema}[S. Martin, \cite{martin}] If $\gamma\in H^\bullet(\mathbb G;\mathbb C)$ admits the lift $\tilde\gamma\in H^\bullet(\Pi;\mathbb C)$, then
\begin{equation}\label{martin}\int_\mathbb G\gamma=\frac{(-1)^{\binom{k}{2}}}{k!}\int_\Pi\tilde\gamma\cup_{\Pi}\Delta^2,
\end{equation}where \[\Delta:=\prod_{1\leq i< j\leq k}(x_i-x_j).\]
\end{teorema}

\begin{cor}[Ellingsrud-Strømme, \cite{ell}] \label{ell-str}The linear morphism
\[\vartheta\colon H^\bullet(\mathbb G;\mathbb C)\to H^\bullet(\Pi;\mathbb C)\colon\gamma\mapsto\tilde\gamma\cup_{\Pi}\Delta
\]is injective, and its image is the subspace of anti-symmetric part of $H^\bullet(\Pi,\mathbb C)$ w.r.t. the $\frak S_r$-action. Moreover
\[\vartheta(\alpha\cup_{\mathbb G}\beta)=\vartheta(\alpha)\cup_{\Pi}\tilde\beta=\tilde\alpha\cup_{\Pi}\vartheta(\beta).
\]
\end{cor}
\proof If $\vartheta(\gamma)=0$, then 
\[\int_\mathbb G\gamma\cup\gamma'=\frac{(-1)^{\binom{k}{2}}}{k!}\int_\Pi\left(\tilde\gamma\cup\Delta\right)\cup\left(\tilde\gamma'\cup\Delta\right)=0
\]for all $\gamma'\in H^\bullet(\mathbb G;\mathbb C)$. Then $\gamma=0$. Being clear that $\vartheta(\gamma)$ is anti-symmetric, observe that any anti-symmetric class is of the form $\tilde\gamma\cup\Delta$ with $\tilde\gamma$ symmetric in $x_1,\dots,x_k$. The last statement follows from the fact that the lift of a cup product is the cup product of the lifts.
\endproof

We can identify the anti-symmetric part of $H^\bullet(\Pi;\mathbb C)\cong H^\bullet(\mathbb P;\mathbb C)^{\otimes k}$ with $\bigwedge\nolimits^kH^\bullet(\mathbb P;\mathbb C)$, using the identifications $i,j$ illustrated in the following \emph{non-commutative} diagram
\[\xymatrix{
H^\bullet(\mathbb P;\mathbb C)^{\otimes k}\ar[r]^{\pi}&\bigwedge\nolimits^kH^\bullet(\mathbb P;\mathbb C)\ar[dl]_{i}\\
[H^\bullet(\mathbb P;\mathbb C)^{\otimes k}]^{\text{ant}}\ar@{^{(}->}[u]\ar@/_1pc/[ur]_{j}&
}\]
where
\[\pi\colon H^\bullet(\mathbb P;\mathbb C)^{\otimes k}\to\bigwedge\nolimits^kH^\bullet(\mathbb P;\mathbb C)\colon\alpha_1\otimes\dots\otimes\alpha_k\mapsto\alpha_1\wedge\dots\wedge\alpha_k,
\]

\[i\colon\bigwedge\nolimits^kH^\bullet(\mathbb P;\mathbb C)\to [H^\bullet(\mathbb P;\mathbb C)^{\otimes k}]^{\text{ant}}\colon\alpha_1\wedge\dots\wedge\alpha_k\mapsto\sum_{\rho\in\frak S_k}\varepsilon(\rho)\alpha_{\rho(1)}\otimes\dots\otimes\alpha_{\rho(k)},
\]together with its inverse
\[j\colon[H^\bullet(\mathbb P;\mathbb C)^{\otimes k}]^{\text{ant}}\to\bigwedge\nolimits^kH^\bullet(\mathbb P;\mathbb C)\colon \alpha_1\otimes \dots\otimes \alpha_k\mapsto \frac{1}{k!}\alpha_1\wedge\dots\wedge\alpha_k.
\]

The Poincaré pairing $g^{\mathbb P}$ on $H^\bullet(\mathbb P;\mathbb C)$ induces a metric $g^{\otimes\mathbb P}$ on $H^\bullet(\mathbb P;\mathbb C)^{\otimes k}$ and a metric $g^{\wedge\mathbb P}$ on $\bigwedge\nolimits^kH^\bullet(\mathbb P;\mathbb C)$ given by
\[g^{\otimes\mathbb P}(\alpha_1\otimes\dots\otimes\alpha_k,\beta_1\otimes\dots\otimes\beta_k):=\prod_{i=1}^kg^{\mathbb P}(\alpha_i,\beta_i),
\]
\[g^{\wedge\mathbb P}(\alpha_1\wedge\dots\wedge\alpha_k,\beta_1\wedge\dots\wedge\beta_k):=\det\left(g^{\mathbb P}(\alpha_i,\beta_j)\right)_{1\leq i,j\leq k}.
\]
Using the identifications above, when $g^{\otimes\mathbb P}$ is restricted on the subspace $[H^\bullet(\mathbb P;\mathbb C)^{\otimes k}]^{\text{ant}}$ it coincides with $k!g^{\wedge\mathbb P}$ on $\bigwedge\nolimits^kH^\bullet(\mathbb P;\mathbb C)$. We deduce the 
\begin{cor}\label{corisometry}
The isomorphism
\[j\circ\vartheta\colon\left( H^\bullet(\mathbb G;\mathbb C), g^{\mathbb G}\right)\to\left(\bigwedge\nolimits^kH^\bullet(\mathbb P;\mathbb C), (-1)^{\binom{k}{2}}g^{\wedge\mathbb P}\right)
\]is an isometry.
\end{cor}
\proof It follows immediately from the integration formula \eqref{martin}.\endproof
An additive basis of $H^\bullet(\mathbb G;\mathbb C)$ is given by the Schubert classes (Poincaré-dual to the Schubert cycles), given in terms of $x_1,\dots, x_k$ by the Schur polynomials
\[\sigma_\lambda:=\frac{\det\begin{pmatrix}
x_1^{\lambda_1+k-1}&x_1^{\lambda_2+k-2} &\dots& x_1^{\lambda_k}\\
x_2^{\lambda_1+k-1}&x_2^{\lambda_2+k-2} &\dots& x_2^{\lambda_k}\\
&&\vdots\\
x_k^{\lambda_1+k-1}&x_k^{\lambda_2+k-2} &\dots& x_k^{\lambda_k}
\end{pmatrix}
}{\det
\begin{pmatrix}
x_1^{k-1}&x_1^{k-2}&\dots&1\\
x_2^{k-1}&x_2^{k-2}&\dots&1\\
&&\vdots\\
x_k^{k-1}&x_k^{k-2}&\dots&1
\end{pmatrix}
}
\]where $\lambda$ is a partition whose corresponding Young diagram is contained in a $k\times (n-k)$ rectangle. The lift of each Schubert class to $H^\bullet(\Pi;\mathbb C)$ is the Schur polynomial in $x_1,\dots, x_k$ (indeed each $x_i$ in the Schur polynomial has exponent at most $n-k<n$). Thus, under the identification above, the class $j\circ\vartheta(\sigma_\lambda)$ is $\sigma_{\lambda_1+k-1}\wedge\dots\wedge\sigma_{\lambda_k}\in\bigwedge\nolimits^kH^\bullet(\mathbb P;\mathbb C)$, $\sigma$ being the generator of $H^2(\mathbb P;\mathbb C)$.

Using the Künneth isomorphism $H^\bullet(\Pi;\mathbb C)\cong H^\bullet(\mathbb P;\mathbb C)^{\otimes k}$, the cup product $\cup_{\Pi}$ is expressed in terms of $\cup_{\mathbb P}$ as follows: 
\[\left(\sum_i\alpha_1^i\otimes\dots\otimes \alpha_k^i\right)\cup_{\Pi}\left(\sum_j\beta_1^j\otimes\dots\otimes \beta_k^j\right)=\sum_{i,j}(\alpha_1^i\cup_{\mathbb P}\beta_1^j)\otimes\dots\otimes(\alpha_k^i\cup_{\mathbb P}\beta_k^j).
\]

If $\gamma\in H^\bullet(\Pi;\mathbb C)^{\frak S_k}$, then $\gamma\cup_{\Pi}(-)\colon H^\bullet(\Pi;\mathbb C)\to H^\bullet(\Pi;\mathbb C)$ leaves invariant the subspace of anti-symmetric classes. Thus, $\gamma\cup_{\Pi}(-)$ induces an endomorphism $A_{\gamma}\in\End\left(\bigwedge\nolimits^kH^\bullet(\mathbb P;\mathbb C)\right)$ that acts on decomposable elements $\alpha=\alpha_1\wedge\dots\wedge\alpha_k$ as follows 
\begin{equation}\label{22-Feb-16-1}A_\gamma(\alpha)=j(\gamma\cup_{\Pi}i(\alpha))=\frac{1}{k!}\sum_{i,\rho}\varepsilon(\rho)(\gamma_1^i\cup_{\mathbb P}\alpha_{\rho(1)})\wedge\dots\wedge(\gamma_k^i\cup_{\mathbb P}\alpha_{\rho(k)}),
\end{equation}
where $\gamma_j^i\in H^\bullet(\mathbb P;\mathbb C)$ are such that
\[\gamma=\sum_{i}\gamma_1^i\otimes\dots\otimes\gamma_k^i.
\]

As an example, in the following Proposition we reformulate  in $\bigwedge\nolimits^kH^\bullet(\mathbb P;\mathbb C)$ the classical Pieri formula, expressing the multiplication by a special Schubert class $\sigma_\ell$ in $H^\bullet(\mathbb G;\mathbb C)$
\[\sigma_\ell\cup_{\mathbb G}\sigma_\mu=\sum_{\nu}\sigma_\nu,
\] where the sum is on all partitions $\nu$ which belong to the set $\mu\otimes\ell$ (the set of partitions obtained by adding $\ell$ boxes to $\mu$, at most one per column) and which are contained in the rectangle $k\times(n-k)$,  in terms of the multiplication by $\sigma_\ell=(\sigma)^\ell\in H^\bullet(\mathbb P;\mathbb C)$. We also make explicit the operation of multiplication by the classes $p_\ell\in H^\bullet(\mathbb G;\mathbb C)$ defined in terms of the special Schubert classes by
\[p_\ell:=-\left(\sum_{\substack{m_1+2m_2+\dots+km_k=\ell \\ m_1,\dots, m_k\geq 0}}\frac{\ell(m_1+\dots+m_k-1)!}{m_1!\dots m_k!}\prod_{i=1}^k(-\sigma_i)^{m_i}\right),\quad \ell=0,\dots, n-1,
\]because of the nice form of their lifts $\tilde p_\ell\in H^\bullet(\Pi;\mathbb C)$.

\begin{prop}\label{23-Feb-16-1}
If $\sigma_\mu\in H^\bullet(\mathbb G;\mathbb C)$ is a Schubert class, then 
\begin{itemize}
\item the product $\sigma_\ell\cup_{\mathbb G}\sigma_\mu$ with a special Schubert class $\sigma_\ell$ is given by
\[j\circ\vartheta(\sigma_\ell\cup_{\mathbb G}\sigma_\mu)=\frac{1}{k!}\left(\sum_{\substack{i_1+\dots+i_k=\ell \\ i_1,\dots, i_k\geq 0}}\sum_{\rho\in\frak S_k}\bigwedge_{j=1}^k\sigma_{i_{\rho(j)}}\cup_{\mathbb P}\sigma_{\mu_j+k-j}\right);
\]
\item the product $p_\ell\cup_{\mathbb G}\sigma_\mu$ is given by
\[j\circ\vartheta(p_\ell\cup_{\mathbb G}\sigma_\mu)=\sum_{i=1}^k\sigma_{\mu_1+k-1}\wedge\dots\wedge(\sigma_{\mu_i+k-i}\cup_{\mathbb P}\sigma_\ell)\wedge\dots\wedge\sigma_{\mu_k}.
\]
\end{itemize}
\end{prop}

\proof From Corollary \eqref{ell-str} we have 
\[\vartheta(\sigma_\ell\cup_{\mathbb G}\sigma_\mu)=\tilde\sigma_\ell\cup_{\Pi}\vartheta(\sigma_\mu)
\]
If $\gamma=\tilde{\sigma}_\ell$ is the lift of the special Schubert class $\sigma_\ell\in H^\bullet(\mathbb G;\mathbb C)$, then 
\[\tilde\sigma_\ell=h_\ell(x_1,\dots, x_k)=\sum_{\substack{i_1+\dots+i_k=\ell \\ i_1,\dots, i_k\geq 0}}\sigma_{i_1}\otimes\dots\otimes\sigma_{i_k},
\]
and using \eqref{22-Feb-16-1} we easily conclude. Analogously, we have that

\[\tilde p_\ell=\sum_{i=1}^kx_i^\ell=\sum_{i=1}^k1\otimes\dots\otimes\underset{\text{$i$-th}}{\sigma_\ell}\otimes\dots\otimes 1,
\]and 
\[A_{\tilde p_\ell}(\alpha)=\sum_{i=1}^k\alpha_1\wedge\dots\wedge(\sigma_\ell\cup_{\mathbb P}\alpha_i)\wedge\dots\wedge\alpha_k.
\]\endproof

\subsection{Quantum Cohomology of $\mathbb G(k,n)$} The identification in the classical cohomology setting of $H^\bullet(\mathbb G;\mathbb C)$ with the wedge product $\bigwedge^k H^\bullet(\mathbb P;\mathbb C)$, exposed in the previous section, has been extended also to the quantum case in \cite{BCFK}, \cite{BCFK2}, \cite{CFKS}, and \cite{KS}.

The following isomorphism of the (small) quantum cohomology algebra of Grassmannians at a point $t\sigma_1=\log q\in H^2(\mathbb G;\mathbb C)$ is well-known
\[QH_q^\bullet(\mathbb G)\cong\frac{\mathbb C[x_1,\dots,x_k]^{\frak S_k}[q]}{\langle h_{n-k+1},\dots,h_n-(-1)^{k-1}q\rangle},
\]while for the (small) quantum cohomology algebra of $\Pi$, being equal to the $k$-fold tensor product of the quantum cohomology algebra of $\mathbb P$, we have 
\[QH_{q_1,\dots,q_k}^\bullet(\Pi)\cong \frac{\mathbb C[x_1,\dots,x_k][q_1,\dots, q_k]}{\langle x_1^n-q_1,\dots, x_k^n-q_k\rangle}.
\]
Following \cite{BCFK}, and interpreting now the parameters $q$'s just as formal parameters, if we denote by $\overline{QH}^\bullet_q(\Pi)$ the quotient of $QH_{q_1,\dots,q_k}^\bullet(\Pi)$ obtained by substituting $q_i=(-1)^{k-1}q$, and denoting the canonical projection by
\[[-]_q\colon QH_{q_1,\dots,q_k}^\bullet(\Pi)\to\overline{QH}^\bullet_q(\Pi),
\]
we can extend by linearity the morphisms $\vartheta,j$ of the previous section to morphisms
\[\overline{\vartheta}\colon QH_q^\bullet(\mathbb G)\to \overline{QH}^\bullet_q(\Pi),
\]
\[\overline{j}\colon \left[\overline{QH}^\bullet_q(\Pi)\right]^{\text{ant}}\to \left(\bigwedge\nolimits^k H^\bullet (\mathbb P;\mathbb C)\right)\otimes_{\mathbb C} C[q].
\]
Notice that the image under $\overline\vartheta$ of any Schubert class $\sigma_\lambda$ is equal to the classical product $\tilde\sigma_\lambda\cup_{\Pi}\Delta$, the exponents of $x_i$'s in the product $\sigma_\lambda(x)\prod_{i<j}(x_i-x_j)$ being less than $n$; as a consequence, the image of $\overline\vartheta$ is equal to the anti-symmetric part w.r.t. the natural $\frak S_k$ action (permuting the $x_i$'s) $$\left[\overline{QH}^\bullet_q(\Pi)\right]^{\text{ant}}\cong [H^\bullet(\Pi;\mathbb C)]^{\text{ant}}\otimes_{\mathbb C}\mathbb C[q].$$ 

The following result, is a quantum generalization of Corollary \ref{ell-str}.
\begin{teorema}[\cite{BCFK}] \label{teoBCFK}For any Schubert classes $\sigma_\lambda,\sigma_\mu\in H^\bullet (\mathbb G;\mathbb C)$ we have
\begin{equation}\label{BCFK1}\overline\vartheta(\sigma_\lambda*_{\mathbb G,q}\sigma_\mu)=\left[\vartheta(\sigma_\mu)*_{\Pi,q_1,\dots,q_k}\tilde \sigma_\lambda\right]_q.
\end{equation}
In particular, using the identification $\overline j$, we have that
\begin{equation}\label{satakeprod}\overline{j}\circ\overline{\vartheta}(\sigma_\mu *_{\mathbb G,q} p_\ell)=\sum_{i=1}^k\sigma_{\mu_1+k-1}\wedge\dots\wedge \sigma_{\mu_i+k-i}*_{\mathbb P,(-1)^{k-1}q} \sigma_\ell\wedge\dots\wedge \sigma_{\mu_k}.\end{equation}
\end{teorema}

\proof The essence of the result \eqref{BCFK1} is the following identity between 3-point Gromov-Witten invariants of genus 0 of the grassmannian $\mathbb G$ and $\Pi$:
\[\langle\sigma_\mu,\sigma_\nu,\sigma_\rho\rangle^{\mathbb G}_{0,3,d\sigma_{1}^\vee}=\frac{(-1)^{\binom{k}{2}}}{k!}\sum_{d_1+\dots+d_k=d}(-1)^{d(k-1)}\langle\sigma_\mu\Delta,\sigma_\nu,\sigma_\rho\Delta\rangle^{\Pi}_{d_1(\sigma^{(1)}_1)^\vee+\dots+d_k(\sigma^{(k)}_1)^\vee},
\]where $d,d_1,\dots,d_k\geq 0$ and $\sigma_1^\vee$ (resp. $(\sigma^{(i)}_1)^\vee$) is the Poincaré dual homology class of $\sigma_1$ (resp. $\sigma^{(i)}_1$). This is easily proved using the Vafa-Intriligator residue formula (see \cite{ber1}).
Equation \eqref{satakeprod} is an immediate consequence of  \eqref{BCFK1}. Notice that, for $\ell=1$, the equation \eqref{satakeprod} can be rewritten in the form
\begin{align*}\overline{j}\circ\overline{\vartheta}(\sigma_\mu *_{\mathbb G,q} \sigma_1)&=\sum_{i=1}^k\sigma_{\mu_1+k-1}\wedge\dots\wedge\underset{\text{$i$-th}}{\sigma_{\mu_i+k-i+1}}\wedge\dots\wedge\sigma_{\mu_k}\\&+(-1)^{k-1}\delta_{n-1,\mu_1+k-1}\sigma_0\wedge\sigma_{\mu_2+k-2}\wedge\dots\wedge\sigma_{\mu_k}.
\end{align*}
The first term coincides with the classical one, whereas the second term is the quantum correction dictated by the Quantum Pieri Formula (see \cite{bert}).
\endproof

The following Proposition is an immediate consequence of the previous result.
\begin{prop}\label{coordcangrass}
At the point $p=t^2\sigma_1\in H^2(\mathbb G;\mathbb C)$ of the small quantum cohomology of $\mathbb G$, the eigenvalues of the operator $$\mathcal U^\mathbb G_p:=c_1(\mathbb G)*_{q}(-)\colon H^\bullet(\mathbb G;\mathbb C)\to H^\bullet(\mathbb G;\mathbb C)$$ are given by the sums
\[u_{i_1}+\dots+u_{i_k},\quad 1\leq i_1<\dots<i_k\leq n,
\]where $u_1,\dots, u_k$ are the eigenvalues of the corresponding operator $\mathcal U^{\mathbb P}$ for projective spaces at the point $\hat p:= t^2\sigma_1+(k-1)\pi i\sigma_1\in H^2(\mathbb P;\mathbb C)$, i.e.
\[\mathcal U^{\mathbb P}_{\hat p}:=c_1(\mathbb P)*_{(-1)^{k-1}q}(-)\colon H^\bullet(\mathbb P;\mathbb C)\to H^\bullet(\mathbb P;\mathbb C).
\]
\end{prop}

\proof
Let $\varphi_1,\dots,\varphi_{n}$ be the idempotents vectors of $QH_{(-1)^{k-1}q}^\bullet(\mathbb P)$ and let $$\mathcal U^{\mathbb P}_{\hat p}(\varphi_i)=u_i\cdot\varphi_i,\quad i=1,\dots, n.$$Then
\[\left(\overline{j}\circ\overline{\vartheta}\right)^{-1}(\varphi_{i_1}\wedge\dots\wedge\varphi_{i_k}),
\]with $1\leq i_1<i_2<\dots<i_k\leq n$, are eigenvectors of $\mathcal U^\mathbb G_p$ with corresponding eigenvalues $u_{i_1}+\dots+u_{i_k}$.
\endproof

\section{Frequency of Coalescence Phenomenon in $QH^\bullet(\mathbb G(k,n))$}
Given $1\leq k< n$, the canonical coordinates of the quantum cohomology of the projective space $\mathbb P^{n-1}_{\mathbb C}$ at the point $\hat p=t^2\sigma_1+(k-1)\pi i \sigma_1$ in the small locus $H^2(\mathbb P^{n-1}_{\mathbb C};\mathbb C)$ are
\begin{equation}\label{coordcpn}u_h=n\exp\left(\frac{t^2+(k-1)\pi i}{n}\right)\zeta_n^{h-1},\quad \zeta_n:=e^{\frac{2\pi i}{n}},\quad h=1,\dots,n.
\end{equation}
Consequently, by Proposition \ref{coordcangrass}, the canonical coordinates of the quantum cohomology of the Grassmannian $\mathbb G(k,n)$ at the point $p=t^2\sigma_1$ are given by the sums
\begin{equation}\label{coordgrass}n\exp\left(\frac{t^2+(k-1)\pi i}{n}\right)\sum_{j=1}^k\zeta_n^{i_j},
\end{equation}for all possible combinations $0\leq i_1<i_2<\dots<i_k\leq n-1$.
This means that, although general results guarantees the semisimplicity of the small quantum cohomology of Grassmannians (see Remark \ref{ossemi}), it may happens that some Dubrovin canonical coordinates coalesce (i.e. the spectrum of the operator $c_1(\mathbb G(k,n))*_p(-)$ is not simple). More precisely, if there is a point $p\in H^2(\mathbb G(k,n);\mathbb C)$ with coalescing canonical coordinates then all points of the small quantum locus have this property. In such a case, we will simply say that the Grassmannian $\mathbb G(k,n)$ is \emph{coalescing}. In this and in the next sections, we want to answer to the following

\begin{quest}
For which $k$ and $n$ the Grassmannian $\mathbb G(k,n)$ is coalescing?
\end{quest}
\begin{quest}
How much frequent is this phenomenon of coalescence among all Grassmannians?
\end{quest}

For the answers we need some preliminary results.

\subsection{Results on vanishing sums of roots of unity} In this section we collect some useful notions and results concerning the problem of vanishing sums, and more general linear relations among roots of unity. The interested reader can find more details and historical remarks in \cite{mann}, \cite{Conway}, \cite{lenstra}, \cite{zannier}, \cite{zannier1} and the references therein. Following \cite{mann} and the survey \cite{lenstra}, we will say that a relation
\begin{equation}\label{vanishing1}\sum_{\nu=1}^ka_\nu z_\nu=0,\quad a_\nu\in\mathbb Q,
\end{equation}
and $z_\nu$'s are roots of unity is \emph{irreducible} if no proper sub-sum vanishes; this means that there is no relation
\[\sum_{\nu=1}^kb_\nu z_\nu=0,\quad\text{with }b_\nu(a_\nu-b_\nu)=0\text{ for all }\nu=1,\dots,k
\]with at least one but all $b_\nu=0$. 

\begin{teorema}[H.B. Mann, \cite{mann}]\label{mann}Let $z_1,\dots,z_r$ be roots of unity, and $a_1,\dots, a_r\in\mathbb N^*$ such that
\[\sum_{i=1}^ra_i z_i=0.
\]Moreover, suppose that such a vanishing relation is \emph{irreducible}. Then, for any $i,j\in\left\{1,\dots, r\right\}$ we have
\[\left(\frac{z_i}{z_j}\right)^m=1,\quad m:=\prod_{\substack{p\text{ prime}\\ p\leq r}}p.
\]
\end{teorema}

Let $G=\langle a\rangle$ be a cyclic group of order $m$, and let $\zeta_m$ be a fixed primitive $m$-th root of unity. There is a well defined natural morphism of ring
\[\phi\colon\mathbb ZG\to\mathbb Z[\zeta_m]\colon a\mapsto\zeta_m,
\]
so that, we have the following identification
\[\ker\phi\equiv\left\{\begin{aligned}
&\mathbb Z\text{-linear relations among}\\
&\text{the $m$-th roots of unity}
\end{aligned}\right\}.
\]
Let us also introduce 
\begin{itemize}
\item the function $\varepsilon_0\colon\mathbb ZG\to\mathbb Z$, defined by
\[\varepsilon_0\left(\sum_{g\in G}x_gg\right):=\operatorname{card}(\left\{g\colon x_g\neq 0\right\});
\]
\item a natural partial ordering on $\mathbb ZG$, by declaring that given two sums
\[x=\sum_{g\in G}x_gg,\ y=\sum_{g\in G}y_gg,\]
we have $x\geq y\text{ if and only if }x_g\geq y_g\text{ for all }g\in G.$
\end{itemize} 
We define $\mathbb NG:=\left\{x\in\mathbb ZG\colon x\geq 0\right\}$. 

\begin{teorema}[T.Y. Lam, K.H. Leung, \cite{lam}]Suppose that $G$ is a cyclic group of order $m=p_1p_2\dots p_r$, with $p_1<p_2<\dots<p_r$ primes and $r\geq 2$. Let $\phi\colon\mathbb ZG\to\mathbb Z[\zeta_m]$ be the natural map, and let $x,y\in\mathbb NG$ such that $\phi(x)=\phi(y)$. If $\varepsilon_0(x)\leq p_1-1$, then we have
\begin{itemize}
\item[(A)] either $y\geq x$,
\item[(B)] or $\varepsilon_0(y)\geq (p_1-\varepsilon_0(x))(p_2-1)$.
\end{itemize}
In case (A), we have $\varepsilon_0(y)\geq\varepsilon_0(x)$, and in case (B) we have $\varepsilon_0(y)>\varepsilon_0(x)$. \end{teorema}

\begin{cor}\label{lam}
In the same hypotheses of the previous Theorem, let us suppose that $\varepsilon_0(x)=\varepsilon_0(y)$. Then $x=y$.
\end{cor}
\proof We necessarily have case (A), and by symmetry of $x$ and $y$, we conclude.\endproof

\begin{defi}
Let $n\geq 2$, and $0\leq k\leq n$. We will say that $n$ is $k$-\emph{balancing} if there exists a combination of integers $1\leq i_1<\dots <i_k\leq n$ such that
\[\zeta_n^{i_1}+\dots+\zeta_n^{i_k}=0,\quad\zeta_n:=e^\frac{2\pi i}{n}.
\]In other words, there are $k$ distinct $n$-roots of unity whose sum is 0.
\end{defi}

\begin{teorema}[G. Sivek, \cite{sivek}]\label{sivek}
If $n=p_1^{\alpha_1}\dots p_r^{\alpha_r}$, with $p_i$'s prime and $\alpha_i>0$, then $n$ is $k$-balancing if and only if 
\[\left\{k,n-k\right\}\subseteq\mathbb Np_1+\dots+\mathbb N p_r.
\]
\end{teorema}

\subsection{Characterization of coalescing Grassmannians}
Using the results exposed above on vanishing sums of roots of unity, we want to study and quantify the occurrence and the frequency of the coalescence of Dubrovin canonical coordinates in small quantum cohomologies of Grassmannians.
Our first aim is to explicitly describe the following sets, defined for $n\geq 2$:
\[\frak A_n:=\left\{h\colon 0< h< n\text{ s.t. }\mathbb G(h,n)\text{ is coalescing}\right\},
\]together with their complements
\[\widetilde{\frak A}_n:=\left\{h\colon 0< h< n\text{ s.t. }\mathbb G(h,n)\text{ is not coalescing}\right\}.
\]

We need some previous Lemmata.

\begin{lemma}\label{lemma1} The following conditions are equivalent 
\begin{itemize}
\item $k\in\frak A_n$;
\item there exist two combinations
\[1\leq i_1<\dots <i_k\leq n\quad\text{and}\quad1\leq j_1<\dots <j_k\leq n,
\]with $i_h\neq j_h$ for at least one $h\in\left\{1,\dots, k\right\}$, such that
\[\zeta_n^{i_1}+\dots+\zeta_n^{i_k}=\zeta_n^{j_1}+\dots+\zeta_n^{j_k},\quad \zeta_n:=e^\frac{2\pi i}{n}.
\]
\end{itemize}
\end{lemma}
\proof It is an immediate consequence of Proposition \ref{coordcangrass}, and formulae \eqref{coordcpn}, \eqref{coordgrass}. \endproof

\begin{lemma}\label{lemma2}$ $
\begin{enumerate}
\item If $n$ is prime, then $\frak A_n=\emptyset$.
\item If $k\in\left\{2,\dots, n-2\right\}$ is such that $k\in\frak A_n$, then $\left\{\min(k,n-k),\dots,\max(k,n-k)\right\}\subseteq\frak A_n$.
\item If $n$ is $k$-balancing (with $2\leq k\leq n-2$), then $k\in\frak A_n$. Thus, if $P_1(n)\leq n-2$, we have $\left\{P_1(n),\dots,n-P_1(n)\right\}\subseteq\frak A_n$.
\end{enumerate}
\end{lemma}

\proof Point (1) follows from Corollary \ref{lam}. For the point (2), notice that given a linear relation as in Lemma \ref{lemma1} with $k$ roots on both l.h.s. and r.h.s. we can obtain a relation with more terms, by adding to both sides the same roots. For point (3), if we have $\zeta_n^{i_1}+\dots+\zeta_n^{i_k}=0$, then also $\zeta_n\cdot(\zeta_n^{i_1}+\dots+\zeta_n^{i_k})=0$, and Lemma \ref{lemma1} applies. The last statement follows from the previous Theorem \ref{sivek} and point (1).
\endproof

\begin{prop}\label{propchiave}
If $P_1(n)\leq n-2$, then $\min\frak A_n=P_1(n)$.
\end{prop}

\proof Let $k:=\min\frak A_n$. We subdivide the proof in several steps.

\begin{itemize}
\item{Step 1.} Let us suppose that $n$ is \emph{squarefree}. By a straightforward application of Corollary \ref{lam}, from an equality like
\[\zeta_n^{i_1}+\dots+\zeta_n^{i_r}=\zeta_n^{j_1}+\dots+\zeta_n^{j_r},
\]and $r<P_1(n)$ we deduce that necessarily $i_h=j_h$ for all $h=1,\dots, r$. Thus $k=P_1(n)$. This proves the Proposition if $n$ is squarefree.

\item{Step 2.} From now on, $n$ is not supposed to be squarefree. We suppose, by contradiction, that $k<P_1(n)$. Because of the minimality condition on $k$, in an equality
\begin{equation}\label{uguaglianzaradici}\zeta_n^{i_1}+\dots+\zeta_n^{i_k}=\zeta_n^{j_1}+\dots+\zeta_n^{j_k},
\end{equation}we have that $i_h\neq j_h$ for all $h=1,\dots, k$. Multiplying, if necessary, by the inverse of one root of unity, we can suppose that one root appearing in \eqref{uguaglianzaradici} is 1. Moreover, we can rewrite equation \eqref{uguaglianzaradici} as a vanishing sum
\begin{equation}\label{ugrad2}
\sum_{i=1}^{2k}\alpha_i z_i=0,\quad\alpha_i\in\left\{-1,+1\right\}
\end{equation}
and where $z_1,\dots,z_{2k}$ are distinct $n$-roots of unity.

\item{Step 3.} We show that the vanishing sum \eqref{ugrad2} is irreducible. Indeed, if we consider the smallest (i.e. with the least number of terms) proper vanishing sub-sum, then it must have at most $k$ addends, otherwise its complement w.r.t. \eqref{ugrad2} would be a vanishing proper sub-sum with less terms. By application of Theorem \ref{mann} to this smallest sub-sum, we deduce that for all roots $z_i$'s appearing in it, we must have
\[\left(\frac{z_i}{z_j}\right)^m=1, \quad m:=\prod_{\substack{p\text{ prime}\\ p\leq k}}p.
\]Under the assumption $k<P_1(n)$, we have that $\gcd(m,n)=1$, and since also $$\left(\frac{z_i}{z_j}\right)^n=1,\quad \text{we deduce }\frac{z_i}{z_j}=1,$$ which is absurd by minimality of $k$. Thus \eqref{ugrad2} is irreducible.

\item{Step 4.} We now show that the order of any roots appearing in \eqref{ugrad2} must be a squarefree number. By application of Theorem \ref{mann}, we know that for all $i,j$ 
\[\left(\frac{z_i}{z_j}\right)^m=1, \quad m:=\prod_{\substack{p\text{ prime}\\ p\leq 2k}}p.
\]Since for one root in \eqref{ugrad2} we have $z_j=1$, we deduce that $z_i^m=1$ for any roots in \eqref{ugrad2}, and that any orders, being divisors of $m$, must be squarefree.
\item{Step 5.} By applying the argument of Step 1, we conclude.\endproof
\end{itemize}

\begin{teorema}\label{main}The complex Grassmannian $\mathbb G(k,n)$ is coalescing if and only if $P_1(n)\leq k\leq n-P_1(n)$. In particular, all Grassmannians of proper subspaces of $\mathbb C^p$, with $p$ prime, are not coalescing.
\end{teorema}

\proof The proof directly follows from Lemma \ref{lemma2} and Proposition \ref{propchiave}. 
\endproof

\subsection{Dirichlet series associated to non-coalescing Grassmannians, and their rareness} Let us now define the sequence 
\[\tilde{\textnormal{\textcyr{l}}}_n:=\operatorname{card}\left(\widetilde{\frak A}_n\right),\quad n\geq 2.
\]Introducing the Dirichlet series 
\[\widetilde{\textnormal{\textcyr{L}}}(s):=\sum_{n=2}^\infty\frac{\tilde{\textnormal{\textcyr{l}}}_n}{n^s},
\]we want deduce information about $(\tilde{\textnormal{\textcyr{l}}}_n)_{n\geq 2}$ studying properties of the generating function $\widetilde{\textnormal{\textcyr{L}}}(s)$.

\begin{teorema}\label{main0.5}
The Dirichlet series $\widetilde{\textnormal{\textcyr{L}}}(s)$ associated to the sequence $(\tilde{\textnormal{\textcyr{l}}}_n)_{n\geq 2}$ is absolutely convergent in the half-plane $\operatorname{Re}(s)> 2$, where it can be represented by the infinite series
\[ \widetilde{\textnormal{\textcyr{L}}}(s)=\sum_{p\textnormal{ prime}}\frac{p-1}{p^s}\left(\frac{2\zeta(s)}{\zeta(s,p-1)}-1\right).
\]The function defined by $\widetilde{\textnormal{\textcyr{L}}}(s)$ can be analytically continued into (the universal cover of) the punctured half-plane $$\left\{s\in\mathbb C\colon \operatorname{Re}(s)>\overline{\sigma}\right\}\setminus\left\{s=\frac{\rho}{k}+1\colon\begin{aligned}
&\rho \text{ pole or zero of }\zeta(s),\\
k&\text{ squarefree positive integer}
\end{aligned}\right\},$$ $$\overline{\sigma}:=\underset{n\to\infty}{\lim\sup}\ \frac{1}{\log n}\cdot \log\left(\sum_{\substack{k\leq n \\k\textnormal{ composite}}}\tilde{\textnormal{\textcyr{l}}}_k\right),\quad 1\leq\overline\sigma\leq\frac{3}{2},$$ having logarithmic singularities at the punctures. In particular, at the point $s=2$ the following asymptotic estimate holds
\begin{equation}\label{asymptotic-tauberian}\widetilde{\textnormal{\textcyr{L}}}(s)=\log\left(\frac{1}{s-2}\right)+O(1),\quad s\to 2,\quad \operatorname{Re}(s)>2.
\end{equation}
\end{teorema}

\proof
Let $\sigma_a$ be the abscissa of (absolute) convergence for $\widetilde{\textnormal{\textcyr{L}}}(s)$. Since $$\inf\left\{\alpha\in\mathbb R\colon \tilde{\textnormal{\textcyr{l}}}_n=O(n^\alpha)\right\}=1,$$ we have $1\leq \sigma_a\leq 2$. Moreover, the sequence $(\tilde{\textnormal{\textcyr{l}}}_n)_{n\geq 2}$ being positive, by a Theorem of Landau (\cite{chandra}, \cite{ten}) the point $s=\sigma_a$ is a singularity for $\widetilde{\textnormal{\textcyr{L}}}(s)$. For $\operatorname{Re}(s)>\sigma_a$, we have (by Theorem \ref{main})
\begin{equation}\label{18-08-16}
\widetilde{\textnormal{\textcyr{L}}}(s)=\sum_{p\text{ prime}}\frac{p-1}{p^s}+\sum_{n\text{ composite}}\frac{2(P_1(n)-1)}{n^s}.
\end{equation}
Note that
\begin{align*}\sum_{n\text{ composite}}\frac{2(P_1(n)-1)}{n^s}&=\sum_{p\text{ prime}}\sum_{\substack{m\geq 2\\ P_1(m)\geq p}}\frac{2(p-1)}{(pm)^s}\\
&=2\sum_{p\text{ prime}}\frac{p-1}{p^s}\left(-1+\sum_{\substack{m\geq 1\\ P_1(m)\geq p}}\frac{1}{m^s}\right)\\
&=2\sum_{p\text{ prime}}\frac{p-1}{p^s}\left(-1+\prod_{\substack{q\text{ prime}\\ q\geq p}}\sum_{k=0}^\infty\frac{1}{q^{ks}}\right)\\
&=2\sum_{p\text{ prime}}\frac{p-1}{p^s}\left(-1+\zeta(s)\prod_{\substack{q\text{ prime}\\ q< p}}\left(1-\frac{1}{q^s}\right)\right).
\end{align*} 
From this and equation \eqref{18-08-16} it follows that
\[ \widetilde{\textnormal{\textcyr{L}}}(s)=\sum_{p\textnormal{ prime}}\frac{p-1}{p^s}\left(\frac{2\zeta(s)}{\zeta(s,p-1)}-1\right).
\]
Since for any $s$ with $\operatorname{Re}(s)>1$ we have $\lim_n\frac{\zeta(s)}{\zeta(s,p_n-1)}=1$, by asymptotic comparison we deduce that the half-plane of absolute convergence of $\widetilde{\textnormal{\textcyr{L}}}(s)$ coincides with the half-plane of $\zeta_P(s-1)-\zeta_P(s)$, hence $\sigma_a=2$ (\cite{froberg}).

The second Dirichlet series in \eqref{18-08-16} defines an holomorphic function in the half-plane of absolute convergence $\operatorname{Re}(s)> \overline\sigma$, where (\cite{hardy-riesz})
$$\overline{\sigma}:=\underset{n\to\infty}{\lim\sup}\ \frac{1}{\log n}\cdot \log\left(\sum_{\substack{k\leq n \\k\textnormal{ composite}}}\tilde{\textnormal{\textcyr{l}}}_k\right).$$
From the elementary and optimal inequality $P_1(n)\leq n^\frac{1}{2}$, valid for any composite number $n$, we deduce that $\frac{1}{2}\leq \overline\sigma\leq\frac{3}{2}$. Thus, the sequence $(\alpha_n)_{n\in\mathbb N}$ defined by \[\alpha_n:=\frac{1}{\log 2n}\cdot \log\left(\sum_{\substack{k\leq 2n \\k\textnormal{ composite}}}\tilde{\textnormal{\textcyr{l}}}_k\right),
\]is bounded: by Bolzano-Weierstrass Theorem, we can extract a subsequence converging to a positive real number $r$ and, by characterization of the superior limit, we necessarily have $r\leq \overline\sigma$. Notice that we have the trivial estimate
\begin{align*}\sum_{\substack{k\leq 2n \\k\textnormal{ composite}}}\tilde{\textnormal{\textcyr{l}}}_k&=\left(\sum_{\substack{4\leq k\leq 2n \\k\textnormal{ even}}}\tilde{\textnormal{\textcyr{l}}}_k\right)+\left(\sum_{\substack{k\leq 2n \\k\textnormal{ odd composite}}}\tilde{\textnormal{\textcyr{l}}}_k\right)>2(n-1),
\end{align*}and we deduce that $1\leq r$. In conclusion, $1\leq\overline\sigma\leq\frac{3}{2}$.

As a consequence, the function $\widetilde{\textnormal{\textcyr{L}}}(s)$ can be extended by analytic continuation at least up to the half-plane $\operatorname{Re}(s)> \overline\sigma$, and it inherits from the function $\zeta_P(s-1)-\zeta_P(s)$ some logarithmic singularities in the strip $\overline\sigma<\operatorname{Re}(s)\leq 2$: they correspond to the points of the form
\[\frac{\rho}{k}+1,\quad 0<\operatorname{Re}(\rho)\leq 1,
\] where $\rho=1$ or $\zeta(\rho)=0$, and $k$ is a squarefree positive integer. This follows from the well known representation 
\[\zeta_P(s)=\sum_{n=1}^\infty\frac{\mu(n)}{n}\log\zeta(ns),
\]$\mu$ being the Möbius arithmetic function (see \cite{gleisher}, \cite{froberg} and \cite{Titch-zeta}).

For $\rho=k=1$, we find again that $s=2$ is a logarithmic singularity for $\widetilde{\textnormal{\textcyr{L}}}(s)$: the asymptotic expansion \eqref{asymptotic-tauberian} follows from \[
\zeta_P(s)=\log\left(\frac{1}{s-1}\right)+O(1), \quad s\to 1,\ \operatorname{Re}(\rho)>1.
\] 
This completes the proof. \endproof

\begin{cor}\label{cor2}
The following statements are equivalent:
\begin{enumerate}
\item \textnormal{(RH)} all non-trivial zeros of the Riemann zeta function $\zeta(s)$ satisfy $\operatorname{Re}(s)=\frac{1}{2}$;
\item the derivative $\widetilde{\textnormal{\textcyr{L}}}'(s)$ extends by analytic continuation to a meromorphic function in the half-plane $\frac{3}{2}<\operatorname{Re}(s)$ with a single pole of oder one at $s=2$. 
\end{enumerate}
\end{cor}

\begin{oss}
The analytic continuation of the function $\widetilde{\textnormal{\textcyr{L}}}(s)$ beyond the line $\operatorname{Re}(s)=\overline\sigma$ is highly influenced by the analytic continuation of the series
\[\sum_{n\text{ composite}}\frac{\tilde{\textnormal{\textcyr{l}}}_n}{n^s}
\]in the strip $1<\operatorname{Re}(s)<\overline\sigma$. In particular, if in this strip it does not have enough logarithmic singularities annihilating those of $\zeta_P(s-1)-\zeta_P(s)$, then the line $\operatorname{Re}(s)=1$ is necessarily a natural boundary for $\widetilde{\textnormal{\textcyr{L}}}(s)$: indeed, the singularities of $\zeta_P(s-1)$ cluster near all points of this line (\cite{landau}). Notice that $s=\overline\sigma$ is necessarily a singularity for $\widetilde{\textnormal{\textcyr{L}}}(s)$, by Landau Theorem.
\end{oss}

\begin{oss}
If we introduce the sequence $\textcyr{l}_n:=\operatorname{card}\left(\frak A_n\right)$, for $n\geq 2$, and the corresponding generating function
\[\textcyr{L}(s):=\sum_{n=2}^\infty\frac{\textcyr{l}_n}{n^s},
\]the following identity holds:
\[\textcyr{L}(s)+\widetilde{\textcyr{L}}(s)=\zeta(s-1)+\zeta(s).
\]In this sense, $\textcyr{L}(s)$ is ``dual'' to $\widetilde{\textcyr{L}}(s)$.
\end{oss}

\begin{cor}\label{cor1}The following asymptotic expansion holds
\[\sum_{k=2}^n\tilde{\textnormal{\textcyr{l}}}_k\sim\frac{1}{2}\frac{n^2}{\log n}.
\]In particular, the non-coalescing Grassmannians are \emph{rare}:
\[\lim_n\frac{2}{n^2-n}\sum_{k=2}^n\tilde{\textnormal{\textcyr{l}}}_k=0.
\]
\end{cor}
\proof Since the function $\widetilde{\textnormal{\textcyr{L}}}(s)$ is holomorphic at all points of the line $\operatorname{Re}(s)=2$ but $s\neq 2$, and the asymptotic expansion \eqref{asymptotic-tauberian} holds, an immediate application of Ikehara-Delange Tauberian Theorem for the case of singularities of mixed-type (involving both monomial and logarithmic terms in their principal parts) for Dirichlet series, gives the result (see \cite{delange} Theorem IV, and \cite{ten} pag. 350).\newline
Another more elementary (and maybe less elegant) proof is the following: from Theorem \ref{main} we have that
\[\sum_{k=2}^n\tilde{\textnormal{\textcyr{l}}}_k=2(1-n)+\pi_0(n)-\pi_1(n)+2\sum_{j=2}^nP_1(j),
\]and recalling the following asymptotic estimates (see \cite{salat}, \cite{anatomy} or \cite{jak})
\[\pi_\alpha(n)\sim\frac{n^{1+\alpha}}{(1+\alpha)\log n},\quad\alpha\geq 0,
\]
\[\sum_{j=2}^nP_1(n)^m\sim \frac{1}{m+1}\frac{n^{m+1}}{\log n},\quad m\geq1,
\]one concludes.
\endproof

\section{Distribution functions of non-coalescing Grassmannians, and equivalent form of the Riemann Hypothesis}
In this section we want to obtain some more fine results about the distribution of these rare not coalescing Grassmannians. Thus, let us introduce the following 
\begin{defi}For all real numbers $x,y\in\mathbb R_{\geq 2}$, with $x\geq y$, define the function
\[\mathscr H(x,y):=\operatorname{card}\left(\left\{n\leq x\colon n\geq2,\ \tilde{\text{\textcyr{l}}}_n>y\right\}\right).
\]In other words, $\mathscr H$ is the cumulative number of vector spaces $\mathbb C^n$, $2\leq n\leq x$, having more than $y$ non-coalescing Grassmannians of proper subspaces. For $x\in\mathbb R_{\geq 4}$ we will define also the restriction
\[\widehat{\mathscr H}(x):=\mathscr H(x,2x^{\frac{1}{2}}).
\]
\end{defi}

In the following result, we describe some analytical properties of the function $\mathscr H$.

\begin{teorema}\label{main1}$ $
\begin{enumerate}
\item For any $\kappa>1$, the following integral representation\footnote{The integral must be interpreted as a Cauchy Principal Value.} holds 
\[\mathscr H(x,y)=\frac{1}{2\pi i}\int_{\Lambda_\kappa}\left[\left(\frac{\zeta(s)}{\zeta\left(s,\frac{y}{2}+1\right)}-1\right)-\zeta_{P,y+1}(s)+\zeta_{P,\frac{y}{2}+1}(s)\right]\frac{x^s}{s}ds, 
\]valid for $x\in\mathbb R_{\geq 2}\setminus\mathbb N,\quad y\in\mathbb R_{\geq 2}$ (with $y\leq x$), and where $\Lambda_\kappa:=\left\{\kappa+it\colon t\in\mathbb R\right\}$ is the line oriented from $t=-\infty$ to $t=+\infty$.

\item For any $\kappa>1$, the following integral representation holds 
\[\mathscr H(x,y)=\frac{1}{2\pi i}\int_{\Lambda_\kappa}\left[\left(\frac{\zeta(s)}{\zeta\left(s,\frac{y}{2}+1\right)}-1\right)x^s+\zeta_P(s)\left(\frac{(y+2)^s}{2^s}-(y+1)^s\right)\right]\frac{ds}{s}, 
\]valid for $x,y\in\mathbb R_{\geq 2}\setminus\mathbb N$ (with $y\leq x$), and where $\Lambda_\kappa:=\left\{\kappa+it\colon t\in\mathbb R\right\}$ is the line oriented from $t=-\infty$ to $t=+\infty$.

\item The following asymptotic estimate holds uniformly in the range $x\geq y\geq 2$
\[\mathscr H(x,y)=\frac{x}{\zeta\left(1,\frac{1}{2}y+1\right)}\left(e^\gamma\omega\left(\frac{\log x}{\log y}\right)+O\left(\frac{1}{\log y}\right)\right)+O\left(\frac{y}{\log y}\right).
\]\end{enumerate}
\end{teorema}

\proof
The crucial observation is the following: if we consider, for fixed $x$ and $y$, the sets
\[\mathcal A:=\left\{n\colon 2\leq n,\ \tilde{\text{\textcyr{l}}}_n>y\right\},\]
\[\mathcal B:= \left\{n\colon 2\leq n,\ 2P_1(n)-2>y\right\},\]
\[\mathcal C:=\left\{p\text{ prime }\colon p-1\leq y,\ 2p-2>y\right\},
\]then we have $\mathcal C\subseteq \mathcal B$ and $\mathcal A\equiv\mathcal B\setminus\mathcal C$. In this way:
\begin{itemize}
\item the Dirichlet series associated to the sequence $\mathbbm 1_{\mathcal A}(n)$ (indicator function of $\mathcal A$) is the difference of the Dirichlet series associated to $\mathbbm 1_{\mathcal B}(n)$ and $\mathbbm 1_{\mathcal C}(n)$. The first one is given by (see e.g. \cite{ten})
\[\frac{\zeta(s)}{\zeta\left(s,\frac{y}{2}+1\right)}-1,
\]while the second one is given by the difference of partial sums
\[\zeta_{P,y+1}(s)-\zeta_{P,\frac{y}{2}+1}(s).
\]An application of Perron Formula for $x$ not integer gives the integral representation (1) of $\sum_{n\leq x}\mathbbm 1_A(n)$.

\item Moreover, we also get the identity
\begin{equation}\label{equality}
\mathscr H(x,y)=\Phi\left(x,\frac{y}{2}+1\right)-\pi_0(y+1)-\pi_0\left(\frac{y}{2}+1\right).
\end{equation}
For $x$ and $y$ not integer, we can apply Perron Formula separately for the three terms:
\[\Phi\left(x,\frac{y}{2}+1\right)=\frac{1}{2\pi i}\int_{\Lambda_\kappa}\left(\frac{\zeta(s)}{\zeta\left(s,\frac{y}{2}+1\right)}-1\right)\frac{x^s}{s}ds,
\]
\[\pi_0(y+1)=\frac{1}{2\pi i}\int_{\Lambda_\kappa}\zeta_P(s)\left(y+1\right)^s\frac{ds}{s},
\]
\[\pi_0\left(\frac{y}{2}+1\right)=\frac{1}{2\pi i}\int_{\Lambda_\kappa}\zeta_P(s)\left(\frac{y}{2}+1\right)^s\frac{ds}{s}.
\]
The sum of the three terms gives the second integral representation (2).

\item Form equation \eqref{equality}, by applying the well known de Bruijn's asymptotic estimate (\cite{bruijn}, \cite{hand-numb}), we obtain the estimate (3).
\end{itemize}
\endproof

\begin{teorema}\label{main2}The function $\widehat{\mathscr H}$ admits the following asymptotic estimate:
\[\widehat{\mathscr H}(x)=\int_{0}^x\frac{dt}{\log t}+O\left(x^\Theta\log x\right),\quad \text{where }\Theta:=\sup\left\{\operatorname{Re}(\rho)\colon \zeta(\rho)=0\right\}.
\]
Hence the following statements are equivalent:
\begin{enumerate}
\item\textnormal{(RH)} all non-trivial zeros of the Riemann zeta function $\zeta(s)$ satisfy $\operatorname{Re}(s)=\frac{1}{2}$;
\item for a sufficiently large $x$, the following (essentially optimal) estimate holds
\begin{equation}\label{riem}\widehat{\mathscr H}(x)=\int_0^x\frac{dt}{\log t}+O\left(x^\frac{1}{2}\log x\right).
\end{equation}
\end{enumerate}
\end{teorema}
\proof
Using the elementary fact that for any composite number $n$ we have $P_1(n)\leq n^\frac{1}{2}$, we obtain the estimate
\[\Phi(x,x^\frac{1}{2})=\pi_0(x)+O\left(x^\frac{1}{2}\right).
\]
Hence, from the equation \eqref{equality} specialized to the case $y=x^\frac{1}{2}$, and by invoking the Prime Number Theorem, we obtain that
\[\widehat{\mathscr H}(x)=\pi_0(x)+O\left(x^\frac{1}{2}\right).
\]It is well known (see e.g. \cite{ten} pag. 271) that
\[\pi_0(x)=\int_{0}^x\frac{dt}{\log t}+O\left(x^\Theta\log x\right),\quad \Theta:=\sup\left\{\operatorname{Re}(\rho)\colon \zeta(\rho)=0\right\}.
\]Since we have $\Theta\geq\frac{1}{2}$ (Hardy proved in 1914 that $\zeta(s)$ has an infinity of zeros on $\operatorname{Re}(s)=\frac{1}{2}$; see \cite{Hardy-zeta}, and also \cite{Titch-zeta} pag. 256), the estimate for $\widehat{\mathscr H}(x)$ follows. The equivalence with RH is evident.
The optimality of the estimate \eqref{riem} (within a factor of $(\log x)^2$) is a consequence of Littlewood's result (\cite{Littlewood-zeta}; see also \cite{hardy-little-zeta} and \cite{mult-num-the}, Chapter 15) on the oscillation for the error terms in the Prime Number Theorem:
\[\pi_0(x)-\int_{0}^x\frac{dt}{\log t}=\Omega_\pm\left(\frac{x^\frac{1}{2}\log\log\log x}{\log x}\right).
\]This completes the proof.
\endproof

\bibliographystyle{alpha}
\bibliography{biblio_prime}

\end{document}